\documentclass[12pt]{article}
\usepackage{amsmath}
\usepackage{amssymb}
\usepackage{graphicx,psfrag,epsf}
\usepackage{enumerate}
\usepackage{natbib}
\usepackage[section]{placeins}
\usepackage{multirow}
\usepackage{booktabs}
\usepackage{url} 
\newcommand{\killpunct}[1]{}
\newcommand{\I}{\mathrm{I}}
\newcommand{\II}{\mathrm{II}}
\newcommand{\blind}{0}

\begin{document}

\def\spacingset#1{\renewcommand{\baselinestretch}%
{#1}\small\normalsize} \spacingset{1}


\if0\blind
{
  \title{\bf Enhancing single-arm phase II trials by inclusion of matched control patients}
  \author{Johannes Krisam \\
    Institute of Medical Biometry and Informatics, University of Heidelberg\\
		INF 130.3, D-69120 Heidelberg, Germany\\
		 mail: krisam@imbi.uni-heidelberg.de (corresponding author)\\
		\\
    and \\
		\\
    Dorothea Weber \\
     Institute of Medical Biometry and Informatics, University of Heidelberg\\
				INF 130.3, D-69120 Heidelberg, Germany\\
		\\
		and \\
		\\
		Richard F. Schlenk \\
    Department of Internal Medicine V and Internal Medicine VI\\
		Heidelberg University Hospital \\
		INF 410, D-69120 Heidelberg, Germany\\
		NCT-Trial Center, German Cancer Research Center, Heidelberg\\
		INF 130.3, D-69120 Heidelberg,  Germany\\
		\\
    and \\
		\\
		  Meinhard Kieser \\
     Institute of Medical Biometry and Informatics, University of Heidelberg\\
				INF 130.3, D-69120 Heidelberg, Germany\\\vspace{1cm}
		}
  \maketitle
} \fi

\if1\blind
{
  \bigskip
  \bigskip
  \bigskip
  \begin{center}
    {\LARGE\bf Enhancing single-arm phase II trials by inclusion of matched control patients}
\end{center}
  \medskip
} \fi

\bigskip
\begin{abstract}
When a novel treatment has successfully passed phase I, different options to design subsequent phase II trials are available. One approach is a single-arm trial, comparing the response rate in the intervention group against a fixed proportion. Another alternative is to conduct a randomized phase II trial, comparing the new treatment with placebo or the current standard. A significant problem arises in both approaches when the investigated patient population is very heterogeneous regarding prognostic  factors. For the situation that a substantial dataset of historical controls exists, we propose an approach to enhance the classic single-arm trial design by including matched control patients. The outcome of the observed study population can be adjusted based on the matched controls with a comparable distribution of known confounders. We propose an adaptive two-stage design with the options of early stopping for futility and recalculation of the sample size taking the matching rate, number of matching partners, and observed treatment effect into account. The performance of the proposed design in terms of type I error rate, power, and expected sample size is investigated via simulation studies based on a hypothetical phase II trial investigating a novel therapy for patients with acute myeloid leukemia. 
\end{abstract}

\noindent%
\vfill

\newpage
\spacingset{1.45} 
\section{Introduction}
\label{sec:Introduction}

Phase II trials play a crucial role in drug development, serving to determine a first estimate of the treatment effect after a successful phase I trial. Based on this estimate, it is decided whether to initiate a subsequent phase III trial. Recently, many development programs, particularly in oncology, failed \citep{Gan2012}, mainly due to lack of efficacy in either phase II or phase III \citep{Harrison2016}. While the reasons for this high percentage are heterogenous, \citet{Gan2010} pointed out that one problem for failed drug development programs might be the inadequacy of the two commonly used standard approaches to design a phase II trial, namely
\begin{enumerate}
	\item  using a single-arm design in which the observed response rate in the intervention group is compared to a fixed proportion based on clinical judgment and historical data, or  
\item conducting a randomized controlled trial comparing the response rate under the novel treatment against placebo or, if available, the current standard.
\end{enumerate}
Single-arm trials ignore patient heterogeneity by design, \, which is, however, likely to arise if there are strong prognostic factors associated with the treatment response. This underlying patient heterogeneity representing an additional source of outcome variability is ignored by the standard single-arm approach, which assumes that the enrolled patients are identical to those in historical studies  \citep{Gan2010}. On the other hand, randomized controlled trials are likely to exhibit an imbalanced distribution of these factors across treatment arms, thus leading to a highly variable treatment effect estimate \citep{Gan2010}. Even though an analysis can be performed by including known confounders into the statistical model, their imbalanced distribution across treatment arms may result in rather unstable results \citep{Redman2007}. This poses a  problem regarding the subsequent go/no-go decision which has to be made based on a highly variable effect estimate.  Hence, neither of the two standard approaches is appropriate when the patient population is quite heterogeneous, which is, however, often the case, e.g. in acute myeloid leukemia (AML) \citep{Schlenk2017} and cross entity approaches \citep{Horak2017}. 

There are currently some discussions in the scientific community how to make an efficient and valid use of so-called "real world data" (RWD) in the context of clinical trials \citep{Murdoch2013,Corrigan2018}. While the randomized controlled trial is not challenged as the gold standard for the demonstration of treatment efficacy in the context of phase III trials \citep{Hudis2015}, the implementation of RWD might be of use to overcome the previously described disadvantages of the standard phase II designs. While approaches allowing to incorporate such historical data into phase II trials have been proposed in the Bayesian literature \citep{Matano2016, Han2017,Normington2020}, this has, to our knowledge, not been done yet in a frequentist design.

One potential source of RWD to enhance phase II trials is the registry data of historical control patients. As an alternative data source, several clinical trial databases are currently emerging, some of which have some kind of restricted access control mechanism, while other databases have an open access policy \citep{Broes2018}. To name one controlled access example, the Vivli database \citep{Bierer2016} hosts the data of 4,700 clinical trials from academia and industry \citep{Vivli2019}, with the option to analyse and share individual participant-level data. One open-access example is the platform Project Data Sphere (PDS) \citep{Bertagnolli2017},  which recently entered a partnership with Merck KGaA \citep{Merck2018}, thereby enabling PDS to include high quality open access data from rare tumor trials, experimental arm data and RWD. 

If a substantial dataset of historical controls exists, we propose to use these data to enhance a classic single-arm trial. This can be done by matching a number of suitable control patients to each intervention patient. The selection of matched historical control patients can practically be achieved by means of propensity score matching. The proposed adaptive two-stage design allows to deal with uncertainties in the planning stage. Moreover, the design enables to stop the trial for futility in case the interim treatment effect estimate does not seem promising and to recalculate the sample size to be enrolled in the second stage of the trial. 

Our manuscript has the following structure. In Section \ref{sec:General}, we outline some general considerations, describe the underlying statistical framework and depict the characteristics of our design in terms of matching procedure, interim analysis, sample size recalculation procedure, and treatment effect estimation. In Section \ref{sec:Futility}, we investigate the performance of the interim futility rule by means of an approximate formula for the probability to continue the trial after the interim analysis. A simulation study motivated by a clinical trial application is conducted in Section \ref{sec:Simulation}, where we investigate the performance characteristics of {standard phase II designs and} the proposed design  in terms of type I error, power, and expected sample size, and also assess the performance of the matching procedure. In Section \ref{sec:Perf_est}, we conduct another simulation study to investigate the performance of the proposed point and interval estimators. We conclude with a discussion in Section \ref{sec:Discussion}.

\section{General considerations and design characteristics}
\label{sec:General}

In the proposed trial design, we assume that the primary outcome $(Y^i)_{i=1,...}$ is a binary variable, e.g. treatment response, with a value of "1" representing a favorable, and a value of "0" representing an unfavorable outcome. Furthermore, there exists a set of $N_{\text{bin}}$ binary confounders $\{X_{B_j}: j=1,2,...,N_{\text{bin}}\}$ following a Bernoulli distribution with $(X_{B_j}^i)_{i=1,...}\sim_{i.i.d.}\text{Ber}(p_j)$ for all $j=1,2,...,N_{\text{bin}}$ and a set of $N_{\text{cont}}$ known continuous confounders $\{X_{C_k}: k=1,2,...,N_{\text{cont}}\}$. Let furthermore $(X_T^i)_{i=1,...}$ denote the variable assigning the treatment status, with $X_T^i=1$ denoting that patient $i$ was treated with the novel treatment $T$ and $X_T^i=0$ denoting that the patient was treated with the control treatment $C$.
It is assumed that the primary outcome is distributed according to a logistic regression model with 
\begin{align*}
\mathrm{Logit}(Y|X_T,(X_{B_j})_{j=1,...,N_{\text{bin}}},(X_{C_j})_{j=1,...,N_{\text{cont}}})
=\beta_0 +\theta X_T +\sum_{j=1}^{N_{\text{bin}}}\beta_{B_j}X_{B_j}+\sum_{k=1}^{N_{\text{cont}}}\beta_{C_k}X_{C_k}{+\varepsilon},
\end{align*}
where $\beta_0$ is a scaling parameter, $\theta$ is the log odds ratio for the treatment effect, $\beta_{B_j}, j=1,...,N_{\text{bin}}$ and $\beta_{C_k}, k=1,...,N_{\text{cont}}$ are the log odds ratios for the respective confounders{, and $\varepsilon=(\varepsilon^i)_{i=1,...}$ denote the residuals for the log odds ratio of each particular patient. It should be noted that we assume one homogenous treatment effect $\theta$ across all patient strata. }

\subsection{Trial design}
\label{sec:Design}

The goal of our proposed trial design is to assess the treatment effect of the novel intervention in terms of the log odds ratio $\theta$ for the primary outcome, while adjusting for relevant known confounders. It is assumed that there is a pool of $n_C$ historical control patients available. In our design, for each enrolled intervention patient, a suitable number of $M$ control patients is about to be drawn from the pool of control patients via a matching approach, which is described in more detail in Subsection \ref{sec:Matching}. Based on the enrolled patients treated with the novel therapy $T$ and the selected historical control patients, the log odds ratio for the primary outcome $\theta$ is estimated. The aim of our trial can either be

\begin{enumerate}[(i)]
	\item a significant result obtained from a standard hypothesis test assessing the null hypothesis $H_0: \theta\leq 0$ against the alternative $H_1: \theta >0$ at a prespecified one-sided  significance level $\alpha$, or
	\item an observed value $\hat{\theta}$ that lies above a pre-defined threshold of clinical relevance $\theta_{\mathrm{cross}}$ on the log odds ratio scale, hence "crossing the threshold" \citep{Eichler2016}. Formally, this is equivalent to obtaining a significant effect for a hypothesis test at a one-sided $\alpha$ of 0.5 assessing the test problem $H_0: \theta-\theta_{\mathrm{cross}}\leq 0$ versus  $H_1: \theta-\theta_{\mathrm{cross}}> 0$.
\end{enumerate}

\begin{figure}[ht]
\caption{Flowchart of the considered adaptive two-stage design}
\begin{center}
\includegraphics[width=10.5cm]{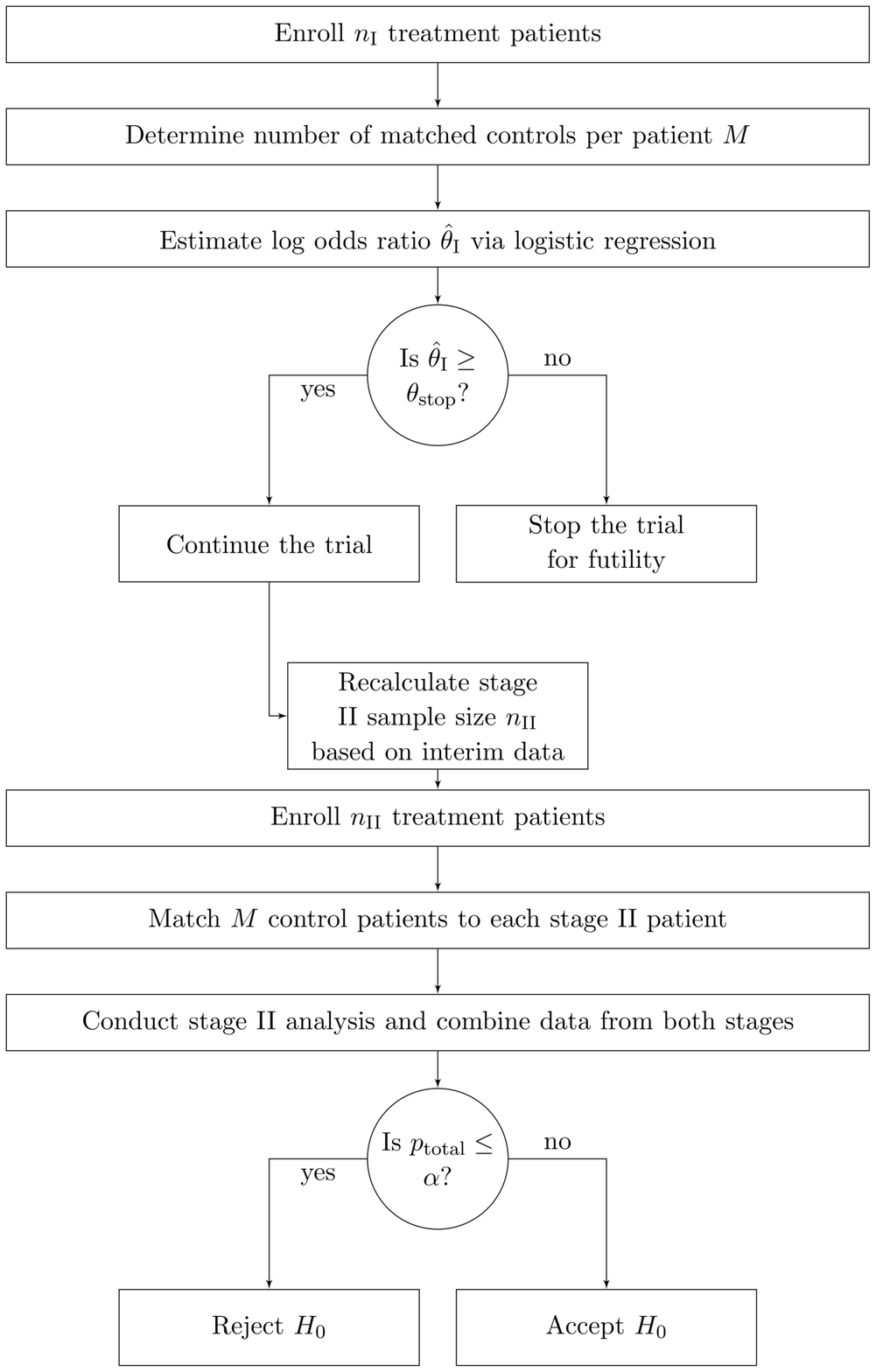}
\end{center}
\label{fig:flowchart} 
\end{figure}

The assessment of the respective hypotheses is done via an adaptive two-stage design, which is depicted by means of a flowchart in Figure \ref{fig:flowchart}. In the proposed design, an initial number of $n_\mathrm{I}$ intervention patients are recruited. After recruitment is completed, a number of $M$ matched control patients is selected for each of the enrolled patients. $M$ is determined using an iterative procedure (see the subsequent Subsection \ref{sec:Matching} for details), together with the matching rate for stage I, $mr_\mathrm{I}$, i.e. the proportion of enrolled patients for which $M$ suitably matching control patients could be found. After the primary outcome is available for all intervention patients to be included in stage I, the interim analysis is conducted. The first step of the interim analysis is the estimation of $\theta$ using a logistic regression model including the matching variables as confounders. In this model, only those intervention patients for whom $M$ matching partners could be found are included. Then, if the estimated log odds ratio $\hat{\theta}_\mathrm{I}$ is smaller than a predefined futility threshold $\theta_\mathrm{stop}$, the trial is stopped for futility and $H_0$ is accepted. Otherwise, the trial will continue to  stage II. Due to the fact that the \textit{a priori} unknown number of matching partners $M$ has an influence on the power of the trial, the sample size for the second trial stage $n_\mathrm{II}$ is recalculated based on the stage I data using a conditional power argument (see Subsection \ref{sec:Interim} for details on the sample size recalculation procedure).

After stage II of the trial has been completed, a 1:$M$ matching is conducted for the stage II intervention patients together with the stage I intervention patients who were not included into the stage I logistic regression model since they did not have $M$ matching partners. Those intervention patients for whom $M$ matching control patients can be found will be included into the stage II logistic regression model together with their matching partners. It should be noted that the controls matched to the stage I patients are not reassigned such that the test statistics of the two stages remain independent in order to ensure type I error rate control. For the final analysis, the data from the two stages need to be combined for the final hypothesis test. In the literature on adaptive designs, there exist several popular approaches to do so, one being the use of a combination test which combines stage-wise \textit{p}-values into an overall \textit{p}-value, e.g. using Fisher's combination test \citep{Bauer1994} or the inverse normal combination test \citep{Lehmacher1999}, the latter of which is used for the proposed trial design. Alternative approaches are the use of the conditional error principle \citep{Proschan1995,Mueller2004} sharing a strong relationship with the combination test methodology \citep{Vandemeulebroecke2006}.

For our design, we make use of the inverse normal method proposed by \citet{Lehmacher1999} and \citet{Cui1999} with \textit{a priori} specified weights $w_\I$ for the stage I and $w_\II$ for stage II, respectively, with $w_\I^2+w_\II^2=1$.  If $\hat{\theta}_\I$ is the log odds ratio estimate from stage I with standard error $\mathrm{se}_\I$, and $\hat{\theta}_\II$ is the log odds ratio estimate from stage II with standard error $\mathrm{se}_\II$, then the respective one-sided \textit{p}-values are defined as
\begin{align*}
p_\I&=1-\Phi\left(\frac{\hat{\theta}_\I-\theta_\mathrm{cross}}{\mathrm{se}_\I}\right),\\
p_\II&=1-\Phi\left(\frac{\hat{\theta}_\II-\theta_\mathrm{cross}}{\mathrm{se}_\II}\right),\text{ and }\\
p_{\mathrm{total}}&=1-\Phi\left(w_\I\Phi^{-1}(1-p_\I)+w_\II \Phi^{-1}(1-p_\II)\right),
\end{align*}
where $\Phi(\cdot)$ denotes the cumulative distribution function of the standard normal distribution and $\Phi^{-1}(\cdot)$ its inverse. Note that in case one conducts a standard hypothesis test and does not use a threshold-crossing approach, $\theta_\mathrm{cross}=0$. If the combined \textit{p}-value $p_{\mathrm{total}}\leq\alpha$, $H_0$ will be rejected. {Since the \textit{p}-values $p_\I$ and $p_\II$ can be assumed as independent and uniformly distributed under the null hypothesis $H_0: \theta-\theta_{\mathrm{cross}}=0$, the same holds true for the combined \textit{p}-value $p_{\mathrm{total}}$, thus ensuring strict and exact control of the type I error rate regardless of the adaptation done at interim \citep{Lehmacher1999}. {While a test using the inverse normal combination method will yield the maximal power when these weights reflect the actual sample sizes of the two stages, the loss in power due to over- or underweighting is rather limited if extreme sample size adaptations are avoided \citep{Lehmacher1999}.  If, e.g. one uses equal weights $w_\I=w_\II=1/\sqrt{2}$ and the ratio of the stage-wise sample sizes fulfills the condition $1/4<n_\II/n_\I<4$, the loss in power is usually less than 3 percent points \citep[p.149]{Wassmer2016}. Since we will later on specify upper and lower bounds for the stage II sample size $n_\II$ that allow to avoid extreme discrepancies between combination test weights and stage-wise sample sizes, choosing equal weights $w_\I=w_\II$ appears to be a sensible choice for our proposed design.}
}

	\subsection{Matching procedure}
	\label{sec:Matching}
	
	Propensity score matching is a method which aims to minimize the effect of observed confounders in estimating treatment effects under the usage of observational data. The propensity score is defined as the conditional probability of the assignment to a specific therapy $T$ under consideration of observed covariates. Hence, the propensity score function $e(X)$ is a balancing function of the observed confounders $X$, such that the conditional distribution of $X$ given the balancing score is the same in both groups and is defined as follows
\begin{align*}
e(X)&= P(T=1 | \boldsymbol{X}),
\end{align*}
under the assumption of
\begin{align*}
P(t_1, \ldots, t_n | X_1, \ldots, X_n)= \prod_{i=1}^N e(X_i)^{t_i} \cdot \bigl( 1-e(X_i)\bigl)^{1-t_i}.
\end{align*}
This propensity score function can be estimated from observed data using, for example, a logistic regression model  \citep{Rosenbaum1983}.

Within the considered adaptive two-stage design, a matching procedure is conducted twice. In stage I, the matching determines the number $M$ of matched controls per patient; furthermore, an extrapolation of the matching rate for recalculating the sample size for the second stage is done. In stage II, the matching is performed to find the fixed number of matching partners $M$ (determined in stage I) for the new patients in stage II. Obviously, a high number of matching partners $M$ will result in a more powerful trial. Nevertheless, the matching rate, which decreases with increasing $M$, should also be sufficiently high, since unmatched intervention patients will not be included into the statistical model, thus lowering the power of the trial. The aim of our interim analysis is thus to determine a suitable number of matched controls $M$ which also ensures an adequately high matching rate.

	The propensity score model is estimated by using baseline characteristics as covariates in a logistic regression model with treatment status as outcome variable. Available research on variable selection in propensity score estimation suggests to include variables that influence the outcome or both the treatment selection and the outcome  \citep{Austin2007}. The resulting pro-pensity scores are transformed to the logit scale  \citep{Rosenbaum1985}. In the described situation, the propensity score model is defined as follows:

\begin{align*}
\hat{e}(X)= \textrm{Logit}\bigl(P(T=1 | \boldsymbol{X}=X)\bigl)=\widetilde{\beta}_0 +\sum_{j=1}^{N_{\text{bin}}}\widetilde{\beta}_{B_j}X_{B_j}+\sum_{k=1}^{N_{\text{cont}}}\widetilde{\beta}_{C_k}X_{C_k}.
\end{align*}

The patients are matched according to the logit of the estimated propensity score to the control patients by using a caliper width of 0.2 of the standard deviation of these estimates  \citep{Austin2007}.

The procedure in stage I starts by conducting a 1:1 propensity score matching and calculating the respective matching rate. Then, a 1:2 propensity score matching is performed and the respective matching rate is calculated. We increase the number of matching partners $M$ as long as the matching rate is equal or higher than the 1:1 matching rate minus a predefined tolerance criterion $\tau$. This parameter defines the maximally allowed deviation from the 1:1 matching rate. If, e.g., $\tau=0$, the 1:2 matching rate should not be smaller than the 1:1 matching rate, otherwise $M$ will be set to 1. Choosing $\tau=0$ represents an approach which ensures that a maximum number of intervention patients is included into the analysis. If, on the other hand, $\tau=0.05$ and the 1:1 matching rate was 0.95, the iterative procedure will increase $M$ as long as the determined 1:$M$ matching rate does not fall below $0.95-0.05=0.9$. The pseudocode for the procedure is shown in Table \ref{SetupMatching}.

\begin{center}
	\begin{table}[h]
		\centering
		\caption{Pseudocode for setup of the matching procedure}
		\begin{tabular}{l}
			\hline
			\textbf{step 1}:
			\vspace{0.2cm}\\
			\quad $M=1$ 
\\
			\quad perform 1:1 propensity score matching
			\\
			\quad calculate matching rate $mr_{1:1}$
		\\
			\quad set $M=2$ \\
			\vspace{0.2cm}\\
			\textbf{step 2}:
		\\
			\quad perform 1:$M$ propensity score matching
		\\
			\quad calculate matching rate $mr_{1:M}$
			\\
			\qquad if ($mr_{1:1} - \tau$ ) $ \leq mr_{1:M}$ \\
			\qquad\quad increase $M$ to $M+1$ and perform step 2\\
			\qquad else \\
			\qquad\quad stop\\
			\hline
		\end{tabular}
		\label{SetupMatching}
	\end{table}

\end{center}
To ensure that $M$ suitable matching partners per intervention patient can still be found in the second step, the maximum number of control patients per intervention patient $M_\mathrm{\max}$ is predefined and the upper limit is given by
\begin{align*}
M_\mathrm{\max}=\left\lfloor\dfrac{\textrm{\small number of patients in the observational data}}{\textrm{\small maximal number of patients in trial}}\right\rfloor.
\end{align*}

After determining the number of matching partners $M$ and the corresponding matching rate, the sample size for stage II is recalculated under consideration of the calculated matching rate, which is described in the following Subsection \ref{sec:Interim}.
The matched pairs in stage I remain unchanged in stage II. Therefore, only the control patients without a matching partner are considered for the matching in stage II. In the analysis of the stage II data, a 1:$M$ propensity score matching is conducted to find $M$ matching partners for the patients enrolled after the interim analysis. Note that the propensity score function in stage II is estimated based on the samples of stage I without a matching partner, the enrolled stage II patients and the remaining controls. The new propensity score model possibly assigns different propensity scores which allows to possibly find a matching partner for unmatched patients of stage I. If this is the case, those patients will be included in the analysis of stage II data together with their matched control patients. Stage I intervention patients for whom $M$ matching control patients could neither be found at the stage I nor at the stage II analysis, as well as stage II intervention patients for whom $M$ matching control patients could not be found at the stage II analysis are not considered in the final analysis.

\subsection{Interim analysis and sample size recalculation}
\label{sec:Interim}

Our proposed trial design includes an interim analysis with the option to stop the trial for futility or to continue to stage II of the trial. Several methods are available for a futility stop within a two-stage design, amongst others stopping rules based on p-values, or criteria involving the conditional power at the interim analysis \citep{Snapinn2006}. As the parameter $\theta$ representing the log odds ratio for the treatment effect plays a key role in our design,  we base the futility decision on a stopping boundary $\theta_{\mathrm{stop}}$ on the log odds ratio scale. This futility stop is considered as non-binding, and hence, no alpha reallocation is conducted. Since the incorporation of a futility stopping rule tends to increase the probability for a type II error, the performance characteristics of the stopping rule need to be assessed in advance. If the threshold for futility stopping $\theta_{\mathrm{stop}}$ is chosen too large, i.e. a very strict stopping rule is applied, this will tend to increase the probability for a trial to be falsely stopped at the interim analysis under $H_1$. A necessary condition for $P[\hat{\theta}\leq\theta_{\mathrm{stop}}|H_1]$ is that this probability should not be higher than the aspired type II probability $\beta$.

If the interim estimate $\hat{\theta}_\I$ successfully crosses the futility threshold $\theta_{\mathrm{stop}}$, the sample size for stage II is recalculated. Several concepts exist for sample size recalculation in adaptive trials, the two most popular approaches being the use of the frequentist conditional power argument \citep{Proschan1995,Lehmacher1999,Liu2001,Friede2001}, which we will also use for our trial design, or the alternative Bayesian concept of predictive power which requires the specification of a prior distribution of the treatment effect \citep{Spiegelhalter1986, Dmitrienko2006,Lan2009}. The option for sample size recalculation has both the advantage to take interim estimates of the treatment effect and parameters which have an influence on the power of the trial like the number of matching partners $M$ and the stage I matching rate $mr_\I$ into account. A crucial aspect is the choice of the treatment effect $\theta_\mathrm{recalc}$ to be used for the sample size recalculation. In this manuscript, we will consider two options for $\theta_\mathrm{recalc}$:

\begin{enumerate}[(i)]
	\item use of the originally assumed log odds ratio $\theta_{\mathrm{plan}}$ to calculate the stage II sample size,
	\item use of the interim estimate $\hat{\theta}_\I$ to calculate the stage II sample size.
\end{enumerate}

While the first option will most likely result in relatively stable sample sizes, the treatment effect assumed in the planning stage might be based on only vague prior information. Hence, the second option has some advantages from a practical point of view, since it allows to incorporate updated knowledge on the treatment effect. Nevertheless, since stage I of the proposed design only incorporates few intervention patients, the interim effect estimate $\hat{\theta}_\I$ might be unstable and thus yield unnecessarily high or too small sample sizes for stage II. The performance of these two rules will later on be investigated by means of a simulation study. Since our trial design incorporates a futility stop, excess sample sizes are unlikely to occur since solely trials with a sufficiently high interim effect estimate continue to stage II. 

Let us assume without loss of generality that $\theta_\mathrm{cross}=0$. In order to calculate the required stage II sample size, we first determine the required information $I_\II$ for the second trial stage in order to ensure a rejection probability at the end of stage II of at least $cp$, the so-called conditional power.  If $\theta\geq\theta_\mathrm{recalc}$, according to the formula by  \citet[p.176]{Wassmer2016}, the stage II information amounts to
\begin{align*}
I_\II&=\frac{\left(\Phi^{-1}(cp)+\Phi^{-1}(1-A(p_\I))\right)^2}{\theta_\mathrm{recalc}^2}
\end{align*}
with
\begin{align*}
A(p_\I)&=1-\Phi\left(\frac{\Phi^{-1}(1-\alpha)-w_\I\Phi^{-1}(1-p_\I)}{w_\II}\right)
\end{align*}
being  the conditional error function for the inverse normal method  \citep[p.153]{Wassmer2016}. It is reasonable to assume that the information $I_\II$ is proportional to $I_\I=1/\mathrm{se}_\I^2$ with respect to the effective sample sizes of the two stages, i.e. $I_\II/I_\I\approx n_\II/(n_\I \cdot mr_\I)$. Hence, we obtain the stage II sample size formula 
\begin{align*}
 \hspace{-0.8cm}n_\II^*=n_\I\cdot mr_\I\cdot  \mathrm{se}_\I^2 \cdot\frac{\left(\Phi^{-1}(cp)+\frac{\Phi^{-1}(1-\alpha)-w_\I\Phi^{-1}(1-p_\I)}{w_\II}\right)^2}{((\theta_\mathrm{recalc}-\theta_\mathrm{cross})^+)^2},
\end{align*}
where $(x)^+$ denotes the so-called positive part of a $x\in\mathbb R$. Note that the denominator of the above fraction needs to be positive since conditional power is defined only for a $\theta_\mathrm{recalc}$ within the parameter space of the alternative hypothesis $H_1$, which may not always be the case when the interim estimate is used for sample size recalculation. 

Regarding the final stage II sample size $\tilde{n}_\II$, a minimum sample size $n_\II^\mathrm{min}$ and a maximum sample size $n_\II^\mathrm{max}$ have to be defined. While $n_\II^\mathrm{max}$ should generally be chosen to reflect available trial resources in terms of trial duration and costs, $n_\II^\mathrm{min}$ should not be too small in order to prevent convergence problems for the stage II logistic regression model. Furthermore, the final stage II sample size $n_\II$ needs to take patients into account for whom $M$ matching partners cannot be found. Hence, an estimate of the stage II matching rate, $\hat{mr}_\II$, needs to be included into the sample size calculation. The determination of $\hat{mr}_\II$ can of course be done using the naive estimate $\hat{mr}_\II=mr_\I$, i.e. the stage I matching rate. We alternatively propose to use the lower limit of the one-sided 99\% Wald-type confidence interval as an estimate for the stage II matching rate, i.e. $\hat{mr}_\II=mr_\I-\Phi^{-1}(0.99)\cdot\sqrt{mr_\I\cdot(1-mr_\I)/(mr_\I\cdot n_\I)}$. Obviously, the second approach leads to a higher stage II sample size. However, it will account for a potentially smaller stage II matching rate due to fewer patients remaining available for matching in the analysis of the stage II data. In the simulation study presented in Section \ref{sec:Simulation}, we will also assess whether the proposed estimation of the stage II matching rate actually provides a more or less accurate estimate of the observed stage II matching rate.

The formula for the actual sample size required to be enrolled in stage II, $\tilde{n}_\II$, is accordingly given by
\begin{align*}
\tilde{n}_\II=\max\left\{n_\II^\mathrm{min},
\min\left\{n_\II^\mathrm{max},n_\II^*/\hat{mr}_\II\right\}\right\}.
\end{align*}

\subsection{Point and interval estimation for treatment effect}
\label{sec:Estimation}

As is the case for all designs with interim adaptations, the sequential and adaptive nature of the trial needs to be taken into account when point and interval estimation of the treatment effect is performed. We will propose some point estimators and interval estimator, and later on investigate the point estimators in terms of bias and root mean square error (RMSE), and the interval estimate in terms of coverage probability.

\subsubsection{Point estimation}

Three different types of point estimates for the log odds ratio of the treatment effect are considered:

\begin{enumerate}[(i)]
	\item the maximum likelihood (ML) estimator 
			\begin{align*}
					\hat{\theta}_\mathrm{ML}=
					\left\{
					\begin{array}{ll}
							\frac{mr_\I n_\I}{mr_\I n_\I+mr_\II \tilde{n}_\II}\hat{\theta}_\I +\frac{\tilde{n}_\II}{mr_\I n_\I+mr_\II \tilde{n}_\II}\hat{\theta}_\II &\, \textrm{ if the trial continues to stage II,} \\							 
							\hat{\theta}_\I & \, \textrm{ else.} \\
					\end{array}\right.
			\end{align*}
	\item a fixed weighted maximum  likelihood (FWML) estimator  \citep{Liu2002}
			\begin{align*}
					\hat{\theta}_\mathrm{FWML}=
					\left\{
					\begin{array}{ll}
						\omega\hat{\theta}_\I +(1-\omega)\hat{\theta}_\II & \, \textrm{ if the trial continues  to stage II,} \\
						\hat{\theta}_\I & \, \textrm{ else, } \\
					\end{array}\right.
			\end{align*}
			with $\omega$ being a pre-specified positive number satisfying $0<\omega<1$. $\omega$ can be chosen deliberately, e.g. as the squared stage I weight used for \textit{p}-value combination, i.e. $\omega=w_\I^2$.
				\item an adaptively weighted maximum  likelihood (AWML) estimator  \citep{Cheng2004}
			\begin{align*}
					\hat{\theta}_\mathrm{AWML}=
					\left\{
					\begin{array}{ll}
						\tilde{\omega}\hat{\theta}_\I +(1-\tilde{\omega})\hat{\theta}_\II  & \, \textrm{ if the trial continues  to stage II,} \\
						\hat{\theta}_\I & \, \textrm{ else, } \\
					\end{array}\right.
					\end{align*}
					with
					\begin{align*}
			\tilde{\omega}=\frac{w_\I/\mathrm{se}_\I}{w_\I/\mathrm{se}_\I+w_\II/\mathrm{se}_\II}
					\end{align*}
					being an adaptive and non-prefixed weight due to its dependence on the stage I and stage II standard errors $\mathrm{se}_\I$ and $\mathrm{se}_\II$.
\end{enumerate}

One should note that using the maximum likelihood estimate will most likely result in some kind of bias due to the dependence between the first-stage effect estimate and the stage II sample size $\tilde{n}_\II$  \citep{Brannath2006}. Hence, alternative point estimates which aim to reduce the bias such as $\hat{\theta}_\mathrm{AWML}$ and $\hat{\theta}_\mathrm{FWML}$ seem to be more promising methods at first hand. The performance of these estimators in terms of bias and root mean square error will be investigated in Section \ref{sec:Perf_est}.

\subsubsection{Interval estimation}

For interval estimation, we propose to use the repeated confidence interval for the inverse normal method  \citep{Wassmer2016}. This one-sided (1-$\alpha)$-confidence interval has the lower bound 
\begin{align*}
	\hat{l}_\mathrm{\theta,1-\alpha}=
					\left\{
					\begin{array}{ll}
							\tilde{\theta}-\frac{\Phi^{-1}(1-\alpha)}{w_\I/\mathrm{se}_\I+w_\II/\mathrm{se}_\II} &\,  \textrm{ if the trial continues to stage II,} \vspace{0mm}\\
							\hat{\theta}_\I-\Phi^{-1}(1-\alpha)\cdot \mathrm{se}_\I & \, \textrm{ else,} \\
					\end{array}\right.
					\end{align*}
where $\tilde{\theta}$ coincides with the adaptively weighted maximum likelihood estimator $\hat{\theta}_\mathrm{AWML}$. In Section \ref{sec:Perf_est}, the coverage probability of this interval estimator will be assessed within the proposed design by means of a simulation study.

\section{Evaluation of the futility rule}
\label{sec:Futility}

As outlined in the previous section, the proposed trial design incorporates the option to stop the trial for futility. Obviously, the futility rule defined by the stopping threshold $\theta_\mathrm{stop}$ should have desirable performance characteristics in the sense that both
\begin{enumerate}[(a)]
	\item  $p_{\mathrm{stop}}^{\theta\in  H_0}=P[\hat{\theta}_\I< \theta_\mathrm{stop} |\theta\leq\theta_\mathrm{cross}]$, being the probability for a futility stop under $H_0$, and
	\item  $p_{\mathrm{continue}}^{ \theta\in  H_1}=P[\hat{\theta}_\I\geq\theta_\mathrm{stop} |\theta>\theta_\mathrm{cross}]$, being the probability for a continuation of the trial after the interim analysis  under $H_1$,
\end{enumerate}

are sufficiently high. Both $p_{\mathrm{stop}}^{\theta\in H_0}$ and  $p_{\mathrm{continue}}^{\theta\in H_1}$ will of course increase with an increasing stage I sample size $n_\I$. While a high stage I sample size is not only beneficial in terms of an increased probability for a correct interim decision but also regarding the sample size recalculation procedure in case the interim estimate is used, $n_\I$ should of course not be chosen too high due to budget and time constraints. Also, a high number of matching partners $M$ will increase the probabilities $p_{\mathrm{stop}}^{\theta\in H_0}$ and $p_{\mathrm{continue}}^{\theta\in H_1}$, since $M$ is proportional to the sample size in the control group. Regarding the choice of the interim decision threshold $\theta_\mathrm{stop}$, the situation is more complicated since a relatively high and thus conservative $\theta_\mathrm{stop}$ will yield a high $p_{\mathrm{stop}}^{\theta\in H_0}$ but a low $p_{\mathrm{continue}}^{\theta\in H_1}$, while the opposite is true for a relatively small and thus liberal $\theta_\mathrm{stop}$. 
When judging the overall design performance in terms of type I and type II error rate control, a small $p_{\mathrm{stop}}^{\theta\in H_0}$ should not affect the type I error rate, since the type I error rate is still controlled by the testing procedure at the significance level $\alpha$. Nonetheless, a small $p_{\mathrm{stop}}^{\theta\in H_0}$ will of course result in a frequent waste of resources and a treatment of more patients with an inactive drug. Not only will trials erroneously continue to stage II with a high probability, but in addition many patients are expected to be enrolled in stage II since the recalculated sample size $\tilde{n}_\II$ is likely to be rather high due to the small treatment effect observed at interim. On the other hand, a high $p_{\mathrm{stop}}^{\theta\in H_1}$ will affect the type II error rate $\beta$ of the trial. This is due to the fact that the probability of a type II error will at least be as high as the probability of a false futility stop, i.e. $\beta\geq p_{\mathrm{stop}}^{\theta\in  H_1}$. 

Hence, when planning a trial with the proposed design, investigations about the performance of the futility stopping rule should be undertaken. While this can of course be done via simulation studies, we will present an analytical method to facilitate application by giving approximate formulae for $p_{\mathrm{stop}}^{H_0}$  and $p_{\mathrm{stop}}^{H_1}$.

We can assume that the interim log odds ratio estimate $\hat{\theta}_\I$ is approximately normally distributed \citep{Agresti1999} and with asymptotic standard error
\begin{align}
\label{eq1}
\begin{split}
\tilde{\mathrm{se}}_\I= \sqrt{\frac{1}{\tilde{n}_\I\cdot\pi_T}+\frac{1}{\tilde{n}_\I\cdot (1-\pi_T)} 
+\frac{1}{\tilde{n}_\I\cdot M\cdot \pi_C}+\frac{1}{\tilde{n}_\I\cdot M \cdot (1-\pi_C)}},
\end{split}
\end{align}
where $\pi_T$ and $\pi_C$ are the proportions of responders amongst the treatment and the control group and $M$ is a fixed number of matching partners. The sample size $\tilde{n}_\I:=mr_\I \cdot n_\I$ is the number of intervention patients for which $M$ suitable controls could be found and can be regarded as the "effective sample size" for stage I.
Hence, 
\begin{align}
\label{eq2}
p_{\mathrm{stop}}^{\theta}\approx  1-\Phi\left(\frac{\theta-\theta_\mathrm{stop}}{\tilde{\mathrm{se}}_\I} \right)&=:\hat{p}_{\mathrm{stop}}^{\theta} \quad \text{and} \quad  p_{\mathrm{continue}}^{\theta}\approx \Phi\left(\frac{\theta-\theta_\mathrm{stop}}{\tilde{\mathrm{se}}_\I} \right)&=:\hat{p}_{\mathrm{continue}}^{\theta} .
\end{align}

Now, as an example for the performance evaluation of the futility threshold, let us consider the following scenarios:

\begin{itemize}
	\item response rates $\pi_T=0.3,\,\pi_C=0.3$ under $H_0$, i.e. $\theta=0$,
	\item response rates $\pi_T=0.5,\,\pi_C=0.3$ under $H_1$, i.e. $\theta=\log(2.33)\approx 0.8473$,
	\item stage I sample sizes $\tilde{n}_\I=10,\,\ldots,\,100$ in steps of 10,
	\item stopping thresholds $\exp(\theta_\mathrm{stop})=1.1,\,1.3,\,1.5$ on the odds ratio scale,
	\item number of matching partners $M=1,\,2,\,5,\,10$.
	
\end{itemize}

\begin{center}
\begin{figure}
\caption{Probabilities to continue to stage II under $H_0(\theta=0$, dashed lines), and under $H_1(\theta=\log(2.33)$, solid lines), in dependence of $\tilde{n}_\I$, denoting the number of stage I intervention patients for which $M$ suitable controls could be found.}
\begin{center}
\includegraphics[width=15cm]{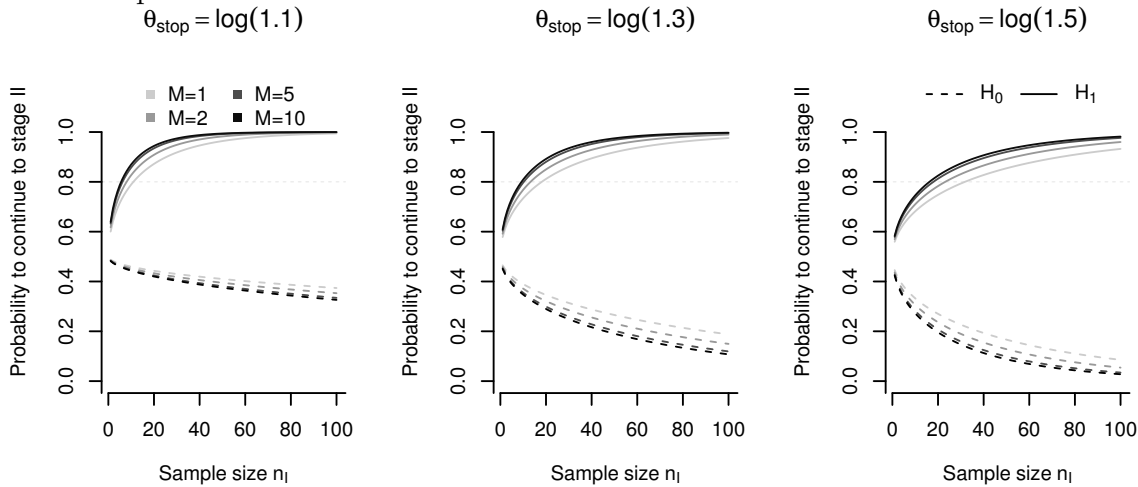}
\end{center}

\label{fig:interimRule}       
\end{figure}
\end{center}

The results for $\hat{p}_{\mathrm{continue}}^{\theta=0}$ and $\hat{p}_{\mathrm{continue}}^{\theta=\log(2.33)}$ are presented in Figure \ref{fig:interimRule}.
Regarding the number of matching partners at the interim analysis, it can be observed that an increasing $M$ will always result in an improved performance of the interim decision rule. Nevertheless, while the performance difference between $M=1$, $M=2$, and $M=5$ is still substantial, the difference between $M=5$ and $M=10$ seems almost negligible. Based on these results, it appears that an extremely high number of matching partners such as $M=10$ is not substantially improving the performance of the interim decision. This can also be observed in the approximation of the standard error $\tilde{\mathrm{se}}_\I$ in equation \ref{eq1}, where an increasing $M$ will only lead to a convergence towards 0 for the last two terms under the root, while the first two terms remain unchanged. 

Obviously, an increasing sample size $n_\I$ results in a better performance of the interim decision rule since the statistical uncertainty about the treatment effect decreases. Regarding the chosen interim decision threshold $\theta_\mathrm{stop}$, we can observe that $\theta_\mathrm{stop}=\log(1.1)$ performs well under the alternative hypothesis $H_1$ such that a probability for correctly continuing to stage II is larger than 0.8 even for small $n_\I$ and small $M$. Recall that in case a power of $1-\beta=0.8$ is to be achieved for the trial, the probability to continue to stage II  under the alternative $p_{\mathrm{continue}}^{\theta\in H_1}$ should at least be 0.8. In case $\theta_\mathrm{stop}=\log(1.3)$, a sample size of at least $\tilde{n}_\I=19$ is required for $p_{\mathrm{continue}}^{\theta\in H_1}\geq 0.8$ if $M=1$. For $\theta_\mathrm{stop}=\log(1.3)$, the required stage I sample size increases to $\tilde{n}_\I=32$ if $M=1$. It should be noted that achieving a value for $p_{\mathrm{continue}}^{\theta\in H_1} \approx 0.8$ is only a necessary requirement to yield a power at least as large as 0.8, since in this case every trial that continues to stage II must terminate with a rejection of $H_0$. This can of course only be achieved with a conditional power $cp$ close to 1, which will result in large sample sizes $\tilde{n}_\II$. Hence, achieving a $p_{\mathrm{continue}}^{\theta\in H_1}$ larger than 0.85 or even 0.9 seems to be more desirable from a practical point of view.
Regarding the performance of the decision rules under $\theta=0$, the most restrictive rule $\theta_\mathrm{stop}=\log(1.5)$ shows the overall best performance, while choosing $\theta_\mathrm{stop}=\log(1.1)$ performs worst. It appears that for $\theta_\mathrm{stop}=\log(1.1)$, increasing the stage I sample size $n_\I$ does not have a substantial influence on $p_{\mathrm{continue}}^{\theta\in H_0}$, which ranges between 0.5 and 0.4, while for the other two decision thresholds, the beneficial effect of an increasing sample size is more pronounced.\\

The previously derived formulae can furthermore be used to determine the conditional power $cp$ applied for sample size recalculation. Since
\begin{align*}
P[\mathrm{reject }H_0|H_1]=p_{\mathrm{continue}}^{\theta\in  H_1}\cdot cp,
\end{align*}

$cp=(1-\beta)/\hat{p}_{\mathrm{continue}}^{\theta_{\mathrm{plan}}}$ seems a sensible choice for the conditional power used for sample size recalculation.  In case that 
$p_{\mathrm{continue}}^{\theta_{\mathrm{plan}}}<(1-\beta)$, a maximum value for conditional power such as $cp=0.99$ can be chosen to recalculate the sample size such that there is a high probability that a trial that was not stopped for futility leads to a rejection of $H_0$ at the final analysis.

It should be noted that the above formulae are only approximate and do not take the additional variance caused by confounders into account. We will hence compare the calculated stopping probabilities with simulated stopping probabilities in the following section.

\section{Simulation study}
\label{sec:Simulation}
Let us consider the situation that a novel salvage treatment for refractory AML patients is to be investigated in a phase II trial. AML is a disease known to be very heterogeneous \citep{Papa2016}. Thus, it can be deemed unlikely that even known predictors of response for refractory AML patients such as age and high risk cytogenetics \citep{Froehling2006} are distributed in a balanced manner across treatment arms when a small randomized phase II trial is conducted \citep{Schlenk2017}. Also, a classical single-arm trial will likely be subject to biased treatment effect estimates in this situation. It is assumed that a large dataset containing AML patients treated with the current standard treatment patients exists. Hence, it would be sensible to conduct a single-arm study with the proposed design in this situation, taking the historical control patients into account.  It is assumed that the response rate under standard salvage therapy amounts to $\pi_C=0.3$, while it is hypothesized that a novel salvage therapy $T$ can increase the response rate to $\pi_T=0.5$.  Based on the historical control data, the prevalence of high risk cytogenetics in both treatment and control patients is assumed to be $0.34$, while age is normally distributed with a mean of 55 years and a standard deviation of 15 years. It is assumed that the response $Y$ in the historical cohort conditional on age and high risk cytogenetics (yes/no) can be modeled using a logistic regression model with
\begin{align*}
\mathrm{Logit}(Y|X_\mathrm{Age},X_\mathrm{Cyto})=&\beta_0+\beta_\mathrm{Age}X_\mathrm{Age}+\beta_\mathrm{Cyto}X_\mathrm{Cyto}{+\varepsilon}.
\end{align*}
Including the variable "treatment" into this model under the assumption of a log odds ratio of $\theta=\log\left(\frac{\pi_T(1-\pi_C)}{\pi_C(1-\pi_T)}\right)$ and assuming no interaction between covariate and treatment results in the logistic model
\begin{align*}
\mathrm{Logit}(Y|X_T,X_\mathrm{Age},X_\mathrm{Cyto})=&\beta_0+\theta X_T+\beta_\mathrm{Age}X_\mathrm{Age}+\beta_\mathrm{Cyto}X_\mathrm{Cyto}{+\varepsilon}.
\end{align*}

Let us assume that fitting this model on the historical control cohort leads to estimated values of $\hat{\beta}_0=2, \hat{\beta}_\mathrm{Age}=-0.05, \hat{\beta}_\mathrm{Cyto}=-0.5$. Since the distribution of response variables is apparently influenced by these two factors, it would be sensible to use these variables for matching historical controls to treatment patients in the proposed design. In the planning stage of the trial, the treatment effect assumed under $H_1$ is $\theta_\mathrm{plan}=\log(0.7/0.3)\approx\log(2.33)$. This underlying model achieves in expectation the envisaged response rates under $H_0$ and $H_1$. {Let us furthermore assume identically independently normally distributed residuals $\varepsilon=(\varepsilon^i)_{i=1,...}\sim_{i.i.d.} N(0,\sigma^2)$.}

In the following simulation study, we consider two situations prior to the start of the trial: the existence of a smaller pool of $n_C=500$ historical control patients, and the existence of a larger pool of $n_C=1000$ historical control patients. In the planning stage of the trial, the stage I sample size of the trial $n_\I$ and the futility stopping threshold $\theta_\mathrm{stop}$ need to be chosen. The methods proposed in Section \ref{sec:Futility} can be used to find suitable parameters which achieve satisfactory probabilities for a correct interim decision under $H_0$ and $H_1$. In the following, we will investigate the case of first-stage sample sizes of $n_\I=20,\,25,$ and 30, and $\theta_\mathrm{stop}=\log(1.3)=0.26$ since the analytical formulae for a correct interim decision which we derived in the previous section indicated a good performance of this threshold under both $H_0$ and $H_1$. Prior to the initiation of a trial with the proposed design, it should additionally be assessed whether sample size recalculation should be based on the treatment effect originally assumed in the planning stage $\theta_\mathrm{plan}$ or the treatment effect estimated at the interim analysis $\hat{\theta}_\I$. This choice is of particular interest when the initially assumed treatment effect may be misspecified. Furthermore, the tolerance criterion $\tau$ used in the algorithm for the determination of the number of matching partners $M$ should be chosen adequately in order to yield both a sufficiently high number of matching partners and high matching rates. 

Also, we conducted simulation studies in which we investigated the performance of the two standard approaches, namely 

\begin{enumerate}
	\item a single-arm trial in which no matched historical controls are incorporated, and which assesses whether the response rate under the new treatment $\pi_T$ exceeds a historical and fixed rate of $0.3$, i.e. in which the hypothesis $H_0: \pi_T\leq 0.3$ is tested against the alternative $H_1: \pi_T>0.3$. Assuming a treatment effect of $\pi_T=0.5$ under $H_1$ would require $n=44$ patients to be enrolled into this single-arm trial when using an approximate version of the binomial test at a one-sided significance level of $\alpha=0.025$ to achieve a power of $1-\beta=0.8$ (calculations performed using ADDPLAN\textsuperscript{TM} v6.1).
	\item a randomized two-arm trial in which no matched historical controls are incorporated, but instead patients are randomly allocated to either receive treatment $T$ or $C$, and which assesses the null hypothesis  $H_0: \pi_T\leq \pi_C$   versus the alternative hypothesis $H_1: \pi_T>\pi_C$. Assuming  $\pi_T=0.5$ and $\pi_C=0.3$ under $H_1$ would require a total sample size of $n=186$ patients to be enrolled into this randomized controlled trial when using the $z$-test for two independent proportions at a one-sided significance level of $\alpha=0.025$. Since this can be deemed as an inconveniently large sample size for a phase II trial, we instead consider the situation that a sample size of $n=100$ patients (50 patients per group) are enrolled due to matters of feasibility (this sample size also represents the chosen upper bound for the simulation study investigating the proposed adaptive design). When using a $z$-test for two proportions at a (liberal) one-sided significance level of $\alpha=0.1$ with this sample size at hand, the power should amount to $1-\beta=0.78$ when assuming that $\pi_T=0.5,\,\pi_C=0.3$ are the underlying response rates. Besides of using the $z$-test, we also consider a scenario in which a logistic regression model adjusting for the prognostic confounders age and high risk cytogenetics was used.
\end{enumerate}

\subsection{Simulation setup}

The performance of the proposed design in terms of type I error rate control, power, and sample size was assessed using a simulation study, which was conducted under the following parameter specifications and assumptions:

\begin{itemize}
	\item significance level $\alpha=0.025$
		\item aspired power $1-\beta=0.8$
	\item stopping threshold $\theta_\mathrm{stop}=\log(1.3)$
	\item crossing threshold $\theta_\mathrm{cross}=0$
	\item assumed treatment effect in the planning stage $\theta_\mathrm{plan}=\log(0.7/0.3)\approx\log(2.33)=0.85$
	\item under $H_0$, a true treatment effect $\theta=0$
	\item under $H_1$,
	\begin{enumerate}
		\item a true treatment effect $\theta=\theta_\mathrm{plan}=\log(0.7/0.3)\approx\log(2.33)=0.85$
		\item a slightly lower treatment effect $\theta=\log(0.7/0.3\cdot 0.48/0.52)\approx\log(2.15)=0.77$
			\item a slightly higher treatment effect $\theta=\log(0.7/0.3\cdot 0.52/0.48)\approx\log(2.53)=0.93$
	\end{enumerate}
	\item to either recalculate the sample size based on $\theta_\mathrm{recalc}=\theta_\mathrm{plan}$, or to recalculate it based on the observed interim effect, i.e. $\theta_\mathrm{recalc}=\hat{\theta}_\I$
	\item conditional power $cp=(1-\beta)/\hat{p}_{\mathrm{continue}}^{\theta_{\mathrm{plan}}}$ used for sample size recalculation, with $\hat{p}_{\mathrm{continue}}^{\theta_{\mathrm{plan}}}$ determined prior to the trial for all possible $M$ using the formulae \ref{eq1} and \ref{eq2}. The value for $M$ determined in the interim analysis then yields the $cp$ chosen to recalculate the sample size. In case $\hat{p}_{\mathrm{continue}}^{\theta_{\mathrm{plan}}}<1-\beta$, $cp$ is set to 0.99.
	\item equal weights for the combination of stage I and II test statistics, i.e. $w_\I^2=w_\II^2=0.5$
		\item stage I sample sizes $n_\I=$ 20, 25, and 30
		\item minimal and maximal stage II sample size of $n_\II^\mathrm{min}=10$ and $n_\II^\mathrm{max}=100-n_\I$
		\item historical control sample sizes $n_C=500$ or $n_C=1000$ 
		\item maximal number of matching partners $M_\mathrm{max}=5$ in case $n_C=500$, and $M_\mathrm{max}=10$ in case $n_C=1000$
		\item matching rate tolerance $\tau=0$, $0.05$, and 0.1.
		\item age in years of control and intervention patients follows i.i.d. normal distributions with expectation 55 and standard deviation 15.
		\item occurrence of high risk cytogenetics of control and intervention patients follows i.i.d. Bernoulli distributions with event probability $p=0.34$.
		\item response $Y^i$ of a given patient is distributed according to the previously described logistic model with $\mathrm{Logit}(Y^i|X^i_\mathrm{Age},X^i_\mathrm{Cyto},X^i_T)=2+\theta X^i_T-0.05 X^i_\mathrm{Age}-0.2 X^i_\mathrm{Cyto}{+\varepsilon^i}$ with residual standard deviations of $\sigma=$ 0, 0.5 and 1.
{		\item For the evaluation of the previously defined outcome distribution under standard phase II designs, $n=44$ patients are enrolled for the single-arm trial scenario, while $n=100$ patients (50 patients per group) are enrolled for the randomized controlled trial scenario. The approximate test for a binomial proportion is used for the single-arm trial at a one-sided significance level of $\alpha=0.025$, and either the $z$-test for two independent proportions or a Wald test from a logistic regression model adjusting for age and high risk cytogenetics is used for the randomized controlled trial, applying a one-sided significance level of $\alpha=0.1$ for both tests.
\item 
}\end{itemize}

100\,000 replications were conducted for each scenario, yielding a standard error of $0.0005  (0.0010)$ for the simulated type I error rate (assuming a true type I error rate of 0.025  (0.1)), and a maximal standard error of $0.0016$ for the simulated power. Furthermore, in all scenarios we used the lower limit of the one-sided 99\% Wald-type confidence interval as an estimate for the stage II matching rate when determining the stage II sample size, as proposed in Subsection \ref{sec:Interim}.

 For the two cases $n_C=500$ and $n_C=1000$, we determined a hypothetical "fixed design" sample size $n_\mathrm{fixed}$ denoting the number of patients required to achieve a power of at least $1-\beta$ under the point alternative of the planned log odds ratio $\theta_\mathrm{plan}=\log(2.33)$ when simply conducting a one-stage study enrolling $n_\mathrm{fixed}$ patients (applying the same matching algorithm and logistic regression model). This was done in order to compare the expected sample size of our procedure with a reference value. For the investigation of the type I error rate, we determined $n_\mathrm{fixed}$ such that a power of at least 0.8 was achieved. For the investigation of the power, we determined $n_\mathrm{fixed}$ such that a power of at least the simulated power of the respective scenario was reached in order to have a fair comparison. It should be noted that this design is only hypothetical since matching rate and number of matching partners $M$ of such a design are not known prior to the start of the trial, and thus no sample size calculation can be performed. The sample size of such a design required to achieve a power of at least $1-\beta$ was also determined using  100\,000 replications per scenario. The same random seed as in the simulations of the adaptive two-stage design was used in order to assess the identical collective of simulated patients. We conducted our simulations using R \citep{R2019} alongside the packages  "Matching" \citep{Sekhon2011} and  "boot" \citep{Davison1997}. All codes used for simulation are supplied as Supplementary Material of this manuscript.

\subsection{Simulation results}

\subsubsection{Type I error rate and power for standard designs}

	\begin{table}
\caption {{Simulated probability to reject the null hypothesis $p_{\mathrm{reject}H_0}$, the average observed response rates in both treatment and control group $\mathrm E[\hat{\pi}_T],\mathrm E[\hat{\pi}_C]$ for a classic single-arm trial  with $n=44$ patients and a randomized controlled trial (RCT) with $n=186$ patients, using $\theta=0$ under $H_0$ and $\theta=\theta_{\mathrm{plan}}=\log(0.7/0.3)$ under $H_1$ for residual standard deviations of $\sigma=0,\,0.5,\,1$.}}
\begin{center}
\begin{tabular}{cccccc}
$\sigma$ & $\theta$ & Model   & $p_{\mathrm{reject}H_0}$ & $\mathrm E[\hat{\pi}_T]$&$\mathrm E[\hat{\pi}_C]$ \\
\toprule
$\sigma=0$ &   $\theta=0$        & single-arm trial 	($\alpha=0.025$)												& 0.0281  & 0.3072  & n.a.     \\
				&						& 	RCT ($z$-test, $\alpha=0.1$)			 & 0.1049   &0.3076 &0.3073    \\
				&						& 	RCT (logistic regression, $\alpha=0.1$)	 & 0.1035 &0.3076 &0.3073   \\

\hline
$\sigma=0$& $\theta=0.85$       & single-arm trial 	($\alpha=0.025$)															& 0.7053  &0.4839 & n.a.  \\
		&								& 	RCT ($z$-test, $\alpha=0.1$)			 			& 0.6963 &0.4839 &0.3073    \\
		&								& 	RCT (logistic regression, $\alpha=0.1$)				 & 0.7410  &0.4839 &0.3073      \\

\hline

$\sigma=0.5$ &   $\theta=0$        & single-arm trial 	($\alpha=0.025$)												& 0.0362  & 0.3144  & n.a.     \\
				&						& 	RCT ($z$-test, $\alpha=0.1$)			 & 0.1045   &0.3143 &0.3143    \\
				&						& 	RCT (logistic regression, $\alpha=0.1$)	 & 0.1036 &0.3143 &0.3143   \\

\hline
$\sigma=0.5$& $\theta=0.85$       & single-arm trial 	($\alpha=0.025$)															& 0.7115  &0.4850 & n.a.  \\
		&								& 	RCT ($z$-test, $\alpha=0.1$)			 			& 0.6678 &0.4843 &0.3143    \\
		&								& 	RCT (logistic regression, $\alpha=0.1$)				 & 0.7154  &0.4843 &0.3143     \\

\hline
$\sigma=1$ &   $\theta=0$        & single-arm trial 	($\alpha=0.025$)												& 0.0595  & 0.3312  & n.a.     \\
				&						& 	 RCT ($z$-test, $\alpha=0.1$)			 & 0.1022   &0.3310 &0.3312    \\
				&						& 	RCT (logistic regression, $\alpha=0.1$)	 & 0.1029 &0.3310 &0.3312   \\

\hline
$\sigma=1$& $\theta=0.85$       & single-arm trial 	($\alpha=0.025$)															& 0.7188  &0.4868 & n.a.  \\
		&								& 	RCT ($z$-test, $\alpha=0.1$)			 			& 0.6064 &0.4859 &0.3312    \\
		&								& 	RCT (logistic regression, $\alpha=0.1$)				 & 0.6468  &0.4859 &0.3312      \\

\bottomrule

\end{tabular}

\end{center}
\label{standard_designs} \end{table}

Table \ref{standard_designs} shows the performance of the classic single-arm and randomized phase II design in terms of the probability to reject $H_0$. 
It can be observed that the standard procedures did generally not achieve the aspired power, and that there was a slightly inflated type I error rate for all designs. In case of the single-arm design, this might be due to the fact that the average response rate for the treatment group under $\theta=0$ observed in our simulations amounted to about 0.307 and thus was slightly larger than the value of 0.3 which was specified as fixed proportion in the assessed null hypothesis. Vice versa, the average observed response rate under $\theta=0.85$ amounted to 0.484, and was thus below the value of $\pi_T=0.5$ assumed in the sample size calculation. Nevertheless, under a binomial test with a true response rate of $\pi_T=0.484$, a single-arm trial using the approximate binomial test with $n=44$ patients should have actually achieved a power of 0.7405 (calculated with ADDPLAN\textsuperscript{TM} v6.1), thus indicating that there was a power loss due to the confounder-induced heterogeneity. An increasing residual standard deviation $\sigma$  resulted in increasing and inflated type I error rates for the single-arm trial, while the power remained more or less constant. In case of the randomized controlled trial, it could be generally observed that the inclusion of the covariates by means of a logistic regression model is a more powerful approach than the use of an unadjusted $z$-test. We identified a minimal inflation of the type I error rate for the randomized controlled trial in all assessed scenarios, which can however be deemed negligible. Nevertheless, it should be noted that using a $z$-test at a significance level of $\alpha=0.1$ with a true response rate of $\pi_T=0.484$ and $\pi_C=0.307$, a randomized controlled trial with $n=100$ patients should actually achieve a power of $0.7045$ (calculated with ADDPLAN\textsuperscript{TM} v6.1), thus indicating a rather small loss in power of only 1  percent point for the $z$-test. For an increasing residual standard deviation $\sigma$, however, the power of the randomized controlled trial decreased substantially.

\subsubsection{Type I error rate, power, and sample size for the proposed two-stage design}

\begin{table}
\caption {Simulated type I error rate (probability to reject the null hypothesis $p_{\mathrm{reject}H_0}$), simulated probability for futility stop $p_{\mathrm{stop}}$, calculated approximate probability for a futility stop $\hat{p}_{\mathrm{stop}}$, expected sample size $\mathrm E[n_\I+\tilde{n}_\II]$,  and hypothetical fixed design sample size $n_\mathrm{fixed}$ required to achieve a power of 0.8 for varying sample sizes $n_\I$ and $n_C$, for $\theta=0$.}
\begin{center}
\begin{tabular}{cccccccc}

$\theta_\mathrm{recalc}$ &$n_C$ & $n_\I$ & $p_{\mathrm{reject}H_0}$ &  $p_{\mathrm{stop}}$ & $\hat{p}_{\mathrm{stop}}$ & $\mathrm E[n_\I+\tilde{n}_\II]$ & $n_\mathrm{fixed}$\\
\hline
$\theta_\mathrm{plan}$& 500 		& 20 & 0.0242    & 0.6807   &   0.6868     &  42.18&  65   \\
										& 	500			& 25 & 0.0240    & 0.7010   &   0.7068     & 43.93 &  65 \\
										& 	500			& 30 & 0.0235    & 0.7189   &   0.7242     & 46.41 &  65 \\
										&   1000	  & 20 &  0.0245   & 0.6931   &   0.6946     &  39.54&  58  \\
										&		1000		&	25 & 0.0252    & 0.7111   &   0.7152     & 41.73 &  58 \\
										&		1000		& 30 & 0.0242    & 0.7288   &   0.7333     & 44.54 &  58 \\

\hline
$\hat{\theta}_\I		$& 500 		& 20 	 & 0.0243 & 0.6807  &   0.6868      & 42.07   &  65 \\
										& 	500			& 25 & 0.0240 & 0.7010  &   0.7068      & 44.69   &  65 \\
										& 	500			& 30 & 0.0235 & 0.7189  &   0.7242      & 47.51   &  65 \\
										&   1000		& 20 & 0.0241 & 0.6931  &   0.6946      & 40.92   &  58 \\
										&		1000		&	25 & 0.0240 & 0.7111  &   0.7152      & 43.82   &  58 \\
										&		1000		& 30 & 0.0242 & 0.7288  &   0.7333      & 46.77   &  58 \\

\hline
\end{tabular}

\end{center}
\small \textbf{Note}: A matching rate tolerance of $\tau=0.05$ and a residual standard
deviation of $\sigma=0$ were used to simulate all scenarios in this table.

\label{H0} \end{table}

 Table \ref{H0} shows the simulation results under the null hypothesis for the proposed two-stage design  in case of a residual standard deviation of $\sigma=0$. It can be observed that the type I error rate is controlled at the defined significance level of $\alpha=0.025$ for all considered scenarios. Also, the observed probabilities for a futility stop are only marginally smaller to those values being calculated based on the formula presented in Section \ref{sec:Futility}, where we inserted the observed average number of matching partners $M$ and the average stage I matching rate $mr_\I$ per simulated scenario. This shows overall good asymptotic properties of our procedure under $H_0$. For the considered situations, the expected sample size lies considerably below the hypothetical fixed design sample size $n_\mathrm{fixed}$ required to achieve a power of 0.8, which was $n_\mathrm{fixed}$=65 for $n_C=500$ and $n_\mathrm{fixed}$=58 for $n_C=1000$. It can be observed that the probability for a correct futility stop is slightly higher in case of a larger pool of control patients, which was to be expected since more control patients are then included into the analysis, resulting in lower standard errors for the interim effect estimator. Similarly, the expected sample size was smaller in case a larger pool of control patients was available. When using the interim estimate for sample size recalculation, the expected sample sizes of the procedure were slightly larger. This might be due to the fact that interim estimates for those trials which were not stopped for futility are still  expected to be smaller than the planned treatment effect under $H_0$, which in turn yielded higher stage II sample sizes. When we investigated the distribution of the sample size, most of its mass was distributed on two data points, being the minimal sample size $n_\I$ and the maximal sample size $100$.  In case of a residual standard deviation of $\sigma=0.5$ and $\sigma=1$, respectively, the type I error rate was maintained as well (results provided in the Appendix).

\begin{table}
\caption {Simulated power (probability to reject the null hypothesis $p_{\mathrm{reject}H_0}$), simulated probability for futility stop $p_{\mathrm{stop}}$, calculated approximate probability for a futility stop $\hat{p}_{\mathrm{stop}}$, expected sample size $\mathrm E[n_\I+\tilde{n}_\II]$,  and hypothetical fixed design sample size $n_\mathrm{fixed}$ required to achieve the observed power in the respective scenario for varying sample sizes $n_\I$ and $n_C$, and for using either $\theta_\mathrm{plan}$ or $\hat{\theta}_\I		$	 for sample size recalculation, for a treatment effect of $\theta=\theta_\mathrm{plan}=\log(2.33)$.}
\begin{center}
\begin{tabular}{cccccccc}

$\theta_\mathrm{recalc}$ &$n_C$ & $n_\I$ & $p_{\mathrm{reject}H_0}$ & $p_{\mathrm{stop}}$ & $\hat{p}_{\mathrm{stop}}$ & $\mathrm E[n_\I+\tilde{n}_\II]$ & $n_\mathrm{fixed}$\\
\hline
$\theta_\mathrm{plan}$					& 500  & 20 				& 0.7736  & 0.1389 &0.1219 			&  61.32  &  60 	\\
										& 	500				& 25 						 	& 0.7839  & 0.1129 &0.0964 			& 59.95 		&  62  \\
										& 	500				& 30 							& 0.7909  & 0.0922 &0.0771 			& 60.31 		&  63  \\
										&   1000			& 20 							& 0.7840  & 0.1262 &0.1099 			&  55.40  &  56  \\
										&		1000		&	25 								& 0.7931  & 0.1009 &0.0851 			& 54.68 		&  57  \\
										&		1000		& 30 								& 0.8038  & 0.0809 &0.0666 			& 55.77		&  59	  \\
                                                                                         
\hline                                                                                   
$\hat{\theta}_\I		$	& 500 & 20 	 									& 0.7415   & 0.1389  &0.1219		& 57.53										&  56  \\
										& 	500			& 25 								& 0.7678 & 0.1129  	&0.0964			& 59.76  									&  60	\\
										& 	500			& 30 								& 0.7861 & 0.0922  	&0.0771			& 62.04 										&  62		 \\
										&   1000		& 20 								& 0.7649 & 0.1262  	&0.1099			& 55.57  									&  53		\\
										&		1000	&	25 									& 0.7904   & 0.1009  &0.0851		& 57.57 										&  57		 \\
										&		1000	& 30 									& 0.8136   & 0.0809  &0.0666		& 59.98  									&  60			 \\

\hline
\end{tabular}\end{center}

\small \textbf{Note}: A matching rate tolerance of $\tau=0.05$ and a residual standard  deviation of $\sigma=0$ were used to simulate all scenarios in this table.
\label{H1} \end{table}

Table \ref{H1} shows the performance of our design under $H_1$ when the treatment effect assumed in the planning stage is actually true, i.e. for $\theta_\mathrm{plan}=\theta=\log(2.33)$. It can be observed that the design achieves the aspired power of about 0.8 for most scenarios. If $n_C=500$ and $n_\I=20$, the power of the design is below 0.75 when using the interim estimate for sample size recalculation. This indicates that the small sample size yielded a relatively unstable interim estimate causing more trials to stop early for futility, which might in turn have caused the loss in power. For the other scenarios, the fact that the aspired power of 0.8 was missed by a narrow margin might be due to the upper bound of 100 patients. This maximal sample size was apparently reached in some cases where the actually recalculated sample size should have been higher in order to yield the aspired power of 0.8. Also, the simulated probabilities to stop for futility are slightly above the calculated ones. This might have caused a further loss in power due to a possibly too small conditional power $cp$. It can be observed that our procedure yielded expected sample sizes which were comparable or even slightly smaller than the sample size required in a fixed design. In general, the higher the number of control and intervention patients, the higher was the probability to reject $H_0$. In case the sample size was recalculated based on the originally assumed treatment effect $\theta_\mathrm{plan}$ and $n_\I=25$ patients were enrolled, this yielded a higher power with smaller expected sample sizes as compared to $n_\I=20$. This indicates a superiority of the former approach in comparison to the latter in this particular scenario. Recalculating the sample size based on the originally assumed treatment effect yielded a slightly higher power with smaller expected sample sizes as compared to using the interim estimate, thus indicating the superiority of the former approach for the scenario $\theta=\theta_\mathrm{plan}=\log(2.33)$. When the interim estimate was used, the expected sample sizes were comparable or slightly above the hypothetical fixed design sample sizes.

\begin{table}
\caption {Simulated power (probability to reject the null hypothesis $p_{\mathrm{reject}H_0}$), probability for futility stop $p_{\mathrm{stop}}$  treatment effect, and expected sample size $\mathrm E[n_\I+\tilde{n}_\II]$ for varying sample sizes $n_\I$ and $n_C$, and for using either $\theta_\mathrm{plan}$ or $\hat{\theta}_\I		$	 for sample size recalculation, for treatment effects $\theta=\log(2.15)$ and $\theta=\log(2.53)$, which differ from the planned effect of $\theta_\mathrm{plan}=\log(2.33)$.}
\begin{center}
\begin{tabular}{ccccccccc}

 & &    &  \multicolumn{3}{c}{$\theta=\log(2.15)$} & \multicolumn{3}{c}{$\theta=\log(2.53)$}   \\
\cmidrule(lr){4-6} \cmidrule(lr){7-9}
$\theta_\mathrm{recalc}$ &$n_C$ & $n_\I$ & $p_{\mathrm{reject}H_0}$ & $p_{\mathrm{stop}}$ & $\mathrm E[n_\I+\tilde{n}_\II]$ & $p_{\mathrm{reject}H_0}$  & $p_{\mathrm{stop}}$ & $\mathrm E[n_\I+\tilde{n}_\II]$  \\
\hline
$\theta_\mathrm{plan}$					& 500  & 20 				  	&  0.7003  &  0.1744 &  61.89  			& 0.8327  &  0.1078 & 60.32 \\		
										& 	500				& 25 						 			&  0.7082 & 0.1486 & 61.02  &  0.8432 & 0.0836 & 58.33 \\
										& 	500				& 30 									&  	0.7163 & 0.1257 & 61.86  & 0.8505 & 0.0650 & 58.41 \\
										&   1000			& 20 							  	& 0.7112  & 0.1616  & 56.31  & 0.8413  & 0.0970  &  54.06 \\
										&		1000		&	25 										& 0.7194 &  0.1352 & 56.02  & 0.8493 &  0.0736 & 53.01 \\
										&		1000		& 30 										&  0.7319 & 0.1127 & 57.48  & 0.8590 & 0.0566 & 53.88 \\

\hline
$\hat{\theta}_\I		$				& 500 			& 20 	 & 0.6701 & 0.1744 & 58.63  & 0.7999 &  0.1078 & 55.99 \\
										& 	500			& 25 					 &  0.6954 & 0.1486 & 61.21  & 0.8250 & 0.0836 &  57.71 \\
										& 	500			& 30 					  & 0.7141 & 0.1257 & 63.98  & 0.8430 & 0.0650 & 59.74 \\
										&   1000		& 20 					     & 0.6967 & 0.1616 & 56.99  & 0.8189 & 0.0970 &  53.73  \\
										&		1000	&	25 						 &   0.7217 & 0.1352 & 59.45  &  0.8429 &  0.0736 &  55.38\\
										&		1000	& 30 						 &  0.7473 & 0.1127 & 62.22  &  0.8626 & 0.0566 &  57.40 \\

\hline
\end{tabular}\end{center}

\small \textbf{Note}: A matching rate tolerance of $\tau=0.05$ and a residual standard deviation of $\sigma=0$ were used to simulate all scenarios in this table.
\label{H1_alt} \end{table}

Table \ref{H1_alt} shows the performance of the design when the treatment effect was misspecified in the planning stage. When we consider  the scenario where the actual treatment effect was smaller than the initially assumed treatment effect ($\theta = \log(2.15)$), using the interim estimate instead of the originally planned treatment effect for sample size recalculation does not yield an advantage in terms of power for $n_C=500$. The approach using $\hat{\theta}_\I$ works slightly better in case  $n_C=1000$ and $n_I\geq 25$, but does not achieve a power close to 0.8. This might be due to the fact that the conditional power used in the simulation was calculated based on the originally assumed treatment effect $\theta_\mathrm{plan}$. In case the actual treatment effect was larger than the originally assumed one ($\theta=\log(2.53)$), using the interim estimate for sample size recalculation showed a better performance in terms of smaller expected sample sizes and a power closer to 0.8, with the exception of the scenario where $n_C=1000$ and $n_\I\geq 25$.

\begin{table}
\caption {{Simulated power (probability to reject the null hypothesis $p_{\mathrm{reject}H_0}$), simulated probability for futility stop $p_{\mathrm{stop}}$, calculated approximate probability for a futility stop $\hat{p}_{\mathrm{stop}}$, expected sample size $\mathrm E[n_\I+\tilde{n}_\II]$,  and hypothetical fixed design sample size $n_\mathrm{fixed}$ required to achieve the observed power in the respective scenario for varying sample sizes $n_\I$ and $n_C$, and for using either $\theta_\mathrm{plan}$ or $\hat{\theta}_\I		$	 for sample size recalculation, for a treatment effect of $\theta=\theta_\mathrm{plan}=\log(2.33)$ and residual standard deviation of $\sigma=0.5$.}}
\begin{center}
\begin{tabular}{cccccccc}

$\theta_\mathrm{recalc}$ &$n_C$ & $n_\I$ & $p_{\mathrm{reject}H_0}$ & $p_{\mathrm{stop}}$ & $\hat{p}_{\mathrm{stop}}$ & $\mathrm E[n_\I+\tilde{n}_\II]$ & $n_\mathrm{fixed}$\\
\hline
$\theta_\mathrm{plan}$					& 500  & 20 				& 0.7406  & 0.1561 &0.1219 			&  61.29  &  61 	\\
										& 	500				& 25 						 	& 0.7501  & 0.1278 &0.0964 			& 60.28 		&  62  \\
										& 	500				& 30 							& 0.7587  & 0.1070 &0.0771 			& 60.71 		&  63  \\
										&   1000			& 20 							& 0.7498  & 0.1432 &0.1099 			&  55.41  &  56  \\
										&		1000		&	25 								& 0.7606  & 0.1164 &0.0851 			& 55.01 		&  57  \\
										&		1000		& 30 								& 0.7709  & 0.0952 &0.0666 			& 56.27		&  59	  \\
                                                                                         
\hline                                                                                   
$\hat{\theta}_\I		$	& 500 & 20 	 									& 0.7105   & 0.1561 &0.1219		& 58.10									&  56  \\
										& 	500			& 25 								& 0.7370 & 0.1278 	&0.0964			& 60.64  									&  60	\\
										& 	500			& 30 								& 0.7584 & 0.1070 	&0.0771			& 62.97										&  63		 \\
										&   1000		& 20 								& 0.7340 & 0.1432  	&0.1099			& 56.13  									&  54		\\
										&		1000	&	25 									& 0.7624   & 0.1164  &0.0851		& 58.53 										&  58		 \\
										&		1000	& 30 									& 0.7853   & 0.0952  &0.0666		& 61.07  									&  61			 \\

\hline
\end{tabular}\end{center}
\label{H1_sigma_0_5} 

\small \textbf{Note}: A matching rate tolerance of $\tau=0.05$ and a residual standard
 deviation of $\sigma=0.5$ were used to simulate all scenarios in this table.
\end{table}

\begin{table}
\caption {{Simulated power (probability to reject the null hypothesis $p_{\mathrm{reject}H_0}$), simulated probability for futility stop $p_{\mathrm{stop}}$, calculated approximate probability for a futility stop $\hat{p}_{\mathrm{stop}}$, expected sample size $\mathrm E[n_\I+\tilde{n}_\II]$,  and hypothetical fixed design sample size $n_\mathrm{fixed}$ required to achieve the observed power in the respective scenario for varying sample sizes $n_\I$ and $n_C$, and for using either $\theta_\mathrm{plan}$ or $\hat{\theta}_\I		$	 for sample size recalculation, for a treatment effect of $\theta=\theta_\mathrm{plan}=\log(2.33)$ and residual standard deviation of $\sigma=1$.}}
\vspace{-0.3cm}
\begin{center}
\begin{tabular}{cccccccc}

$\theta_\mathrm{recalc}$ &$n_C$ & $n_\I$ & $p_{\mathrm{reject}H_0}$ & $p_{\mathrm{stop}}$ & $\hat{p}_{\mathrm{stop}}$ & $\mathrm E[n_\I+\tilde{n}_\II]$ & $n_\mathrm{fixed}$\\
\hline
$\theta_\mathrm{plan}$					& 500  & 20 				& 0.6518  & 0.1974 &0.1219 			&  60.76  &  61 	\\
										& 	500				& 25 						 	& 0.6599  & 0.1699 &0.0964 			& 60.33 		&  63  \\
										& 	500				& 30 							& 0.6678  & 0.1489 &0.0771 			& 61.22 		&  64  \\
										&   1000			& 20 							& 0.6640  & 0.1870 &0.1099 			&  55.17  &  57  \\
										&		1000		&	25 								& 0.6692  & 0.1591 &0.0851 			& 55.34 		&  58  \\
										&		1000		& 30 								& 0.6798  & 0.1370 &0.0666 			& 56.92		&  60	  \\
                                                                                         
\hline                                                                                   
$\hat{\theta}_\I		$	& 500 & 20 	 									& 0.6270   & 0.1974  &0.1219		& 58.90										&  58  \\
										& 	500			& 25 								& 0.6522 &   0.1699  	&0.0964			& 61.86 									&  62	\\
										& 	500			& 30 								& 0.6745 &   0.1489  	&0.0771			& 64.57 										&  65		 \\
										&   1000		& 20 								& 0.6527 &   0.1870  	&0.1099			& 57.27 									&  56		\\
										&		1000	&	25 									& 0.6802   & 0.1591  &0.0851		& 60.11 										&  60		 \\
										&		1000	& 30 									& 0.7039   & 0.1370  &0.0666		& 62.98  									&  63			 \\

\hline

\end{tabular}\end{center}
\label{H1_sigma_1}

\small \textbf{Note}: A matching rate tolerance of $\tau=0.05$ and a residual standard 
deviation of $\sigma=1$ were used to simulate all scenarios in this table.

 \end{table}

Tables \ref{H1_sigma_0_5} and \ref{H1_sigma_1} show the performance of the design under $H_1$ in case of additional variation induced by a  residual with standard deviation $\sigma=0.5$ and $\sigma=1$, respectively. It can be observed in Table \ref{H1_sigma_0_5} that a moderate residual standard deviation of $\sigma=0.5$ leads to a small loss in power, an issue which was also observed earlier for the randomized controlled trial design displayed in Table \ref{standard_designs}, and to a higher probability for falsely stopping the trial at interim. Nevertheless, the expected sample size is still smaller for the proposed design when compared to a corresponding sample size $n_\mathrm{fixed}$ in case one uses the originally assumed effect size for sample size recalculation, indicating that the proposed approach seems to perform well nonetheless. The method using the interim estimate shows a slightly worse performance in comparison, thus reflecting the results that were obtained in the scenarios of a residual standard deviation of $\sigma=0$ (displayed in Table \ref{H1}).  In case of a large residual  standard deviation $\sigma=1$, the power  decreases substantially to values only ranging between 63\% and 70\%. Also, the probability for a false stop for futility increases remarkably if $\sigma=1$.
\subsubsection{Matching procedure}

We also investigated the performance of the matching procedure in terms of the distribution of the number of matching partners $M$, the matching rates in stage I and stage II, $mr_\I$ and $mr_\II$, and the potential influence on power, probability for a futility stop, and expected sample size. For this assessment, we varied our tolerance criterion $\tau$ from the matching algorithm by applying values of $\tau=0,\,0.05,\,0.1$. We only considered the situation of the point alternative $\theta=\theta_\mathrm{plan}=\log(2.33)$ in this investigation. Also, we only investigated scenarios in which the sample size was recalculated using the originally assumed treatment effect $\theta_\mathrm{plan}$.  The results are shown in Table \ref{Matching_Table}.

\begin{table}
\caption {Expected number of matching partners $\mathrm E [M]$, expected matching rates in stage I and stage II $\mathrm E [mr_\I]$ and $\mathrm E [mr_\II]$, and mean estimated stage II matching rate for sample size recalculation, alongside power (probability to reject the null hypothesis $p_{\mathrm{reject}H_0}$), probability for futility stop $p_{\mathrm{stop}}$, and expected sample size $\mathrm E[n_\I+\tilde{n}_\II]$ for a treatment effect of $\theta=\theta_\mathrm{plan}=\log(2.33)$.}
\vspace{-0.3cm}
\begin{center}
\begin{tabular}{cccccccccc}

$\tau$ &$n_C$ & $n_\I$ & $\mathrm E [M]$  & $\mathrm E [mr_\I]$ & $\mathrm E [mr_\II]$ & $\mathrm E [\hat{mr}_\II]$ & $p_{\mathrm{reject}H_0}$  & $p_{\mathrm{stop}}$ & $\mathrm E[n_\I+\tilde{n}_\II]$ \\
\hline
$0$					& 500  				& 20 															&  	4.43  &  0.9977	 &  0.9406		& 0.9728  & 0.7658  & 0.1428 &  62.72   \\
									& 500  				& 25 												&  	4.28  &  0.9977	 &  0.9467		& 0.9754 & 0.7740  & 0.1181 &  61.97   \\
											& 500  				& 30 												&   4.14  &  0.9977	 &  0.9496	&  0.9774 & 0.7815  &  0.0982 &   62.66   \\
									& 1000  				& 20 											&  	8.81  &  0.9989	 &  0.9546		& 0.9817 & 0.7774  & 0.1294 &  56.24   \\
									& 1000  				& 25 											&  	8.52  &  0.9989	 &  0.9586		& 0.9835 & 0.7860  & 0.1043 &  55.94   \\
											& 1000  				& 30 									&  	 8.24 &   0.9989 &  0.9594		  & 0.9848 & 0.7961  &  0.0845 &   57.28   \\
\hline                                                                                           
$0.05$					& 500  						& 20 											&  	4.93  &  0.9862	 &  0.9244		& 0.9255 & 0.7736  & 0.1389 &  61.32   \\
										& 	500				& 25 											&  	4.88  & 0.9864   &  0.9281 		& 0.9325 & 0.7839 & 0.1129 & 59.95 		\\
										& 	500				& 30 											&   4.83 &  0.9866   &  0.9277  	& 0.9378 & 0.7909 & 0.0922 & 60.31 		\\
										&   1000			& 20 			  							&  	9.85  & 0.9866   & 0.9399  		& 0.9268 & 0.7840  & 0.1262  &  55.40  \\
										&		1000			&	25 												& 	9.76 &  0.9868 &  0.9413  	& 0.9337 & 0.7931 &  0.1009 & 54.68 		\\
										&		1000			& 30 												&   9.65 &  0.9870 & 0.9386  		& 0.9389 & 0.8038 & 0.0809 & 55.77			\\
\hline                                                                                           
$0.1$					& 500  				& 20											&  	4.99       &   0.9839	&  0.9215		    &   0.9184 &  0.7739 & 0.1388 &   61.20   \\
							& 500  				& 25 											&  	4.98         &  0.9834	&  0.9237	  & 0.9240 & 0.7846 & 0.1125 &  59.75   \\
							& 500  				& 30 											&  4.99        &  0.9820	&   0.9198	    &  0.9255 & 0.7922 & 0.0915 &   59.98   \\
							& 1000  				& 20 											&  	9.98  &  0.9842	    &  0.9372		  & 0.9193 & 0.7842 & 0.1262 &  55.38   \\
							& 1000  				& 25 											&  	 9.97    &   0.9837	&  0.9372		  &  0.9248 & 0.7939 & 0.1006 &   54.62   \\
							& 1000  				& 30 											&  	9.99        &  0.9823	&   0.9311		  & 0.9263 & 0.8044 &  0.0803 &   55.68   \\
\hline

\end{tabular}

\small \textbf{Note}: For sample size recalculation, the originally assumed treatment effect $\theta_\mathrm{plan}=\log(2.33)$ and a residual standard deviation of $\sigma=0$ were used  to simulate all scenarios in this table.
\end{center}

\label{Matching_Table} \end{table}

It can be observed that the expected number of matching partners $\mathrm E [M]$ decreases with an increasing sample size $n_\I$, since more adequate control patients have to be drawn from the historical dataset. The expected stage II matching rates were smaller than their stage I counterparts. This is plausible since $M$ is not re-determined in the analysis of the stage II data, and suitable partners for the second-stage patients are drawn from a diminished pool of control patients since matching partners from stage I are not reallocated. 

With an increasing tolerance criterion $\tau$, the expected number of matching partners increased while the matching rates for stage I and stage II decreased. While it seems desirable to enroll a high number of matching partners per intervention patient, the matching rate should also be sufficiently high. In order to assess which tolerance criterion should be preferable, it is sensible to take a look at the power and expected sample sizes. Choosing $\tau=0$ yielded a slightly decreased power as compared to $\tau=0.05$, while the expected sample size was even higher. Hence, choosing $\tau=0$ appears to be too strict by allowing no deviation from 1:1 matching rate, thus not optimally making use of the available historical controls.

Our proposed estimator for the stage II matching rate $\hat{mr}_\II$ seems to perform well when $\tau=0.05$ and $\tau=0.1$, approximately yielding the observed matching rates. However, it slightly overestimates the observed stage II matching rate in case of the strict matching tolerance criterion $\tau=0$. This also might be a reason why the aspired power of 0.8 was not reached in these scenarios.

It should be mentioned that in some very few simulation runs, the logistic regression model could not be fitted due to perfect separation of responders and non-responders in the stage II analysis. In these rare situations, the stage II data was ignored for hypothesis testing and estimation of the treatment effect. This aberration generally occurred for less than $0.01\%$ of all simulation runs and can thus be deemed irrelevant regarding the observed overall performance characteristics of our design. Increasing the minimal stage II sample size $n_\II^\mathrm{min}$ will of course reduce the probability for such occurrences. We hence propose not to choose $n_\II^\mathrm{min}$ smaller than 10 patients in order to prevent such situations when conducting a trial according to our design. 

\section{Performance of point and interval estimators}

\label{sec:Perf_est}
In order to assess the performance of the three point estimators and the interval estimator presented in Subsection \ref{sec:Estimation}, we conducted a second simulation study for treatment effects varying the log odds ratio of $\theta=-0.1$ to $\theta=2$ in steps of 0.1. The significance level was set to $\alpha=0.025$ {and the residual standard deviation was $\sigma=0$}. The stage I sample size was set to $n_\I=25$, minimal and maximal stage II sample sizes were $n_\II^\mathrm{min}=10$ and $n_\II^\mathrm{max}=75$, number of control patients was set to $n_C=1000$, sample size recalculation was based on the originally assumed true effect, and equal weights were used for the two stages. In order to assess the performance of the estimators combining two trial stages, no futility stop was incorporated into the design, i.e. $\theta_\mathrm{stop}$ was set to $-\infty$.

\begin{figure}
\caption{Simulated bias and Root Mean Square Error (RMSE) for maximum likelihood (ML), fixed weighted maximum likelihood (FWML) and adaptively weighted maximum likelihood (AWML) estimator of the log odds ratio. Bias and RMSE are displayed on the log odds ratio scale.}
\begin{center}
\includegraphics[width=9cm]{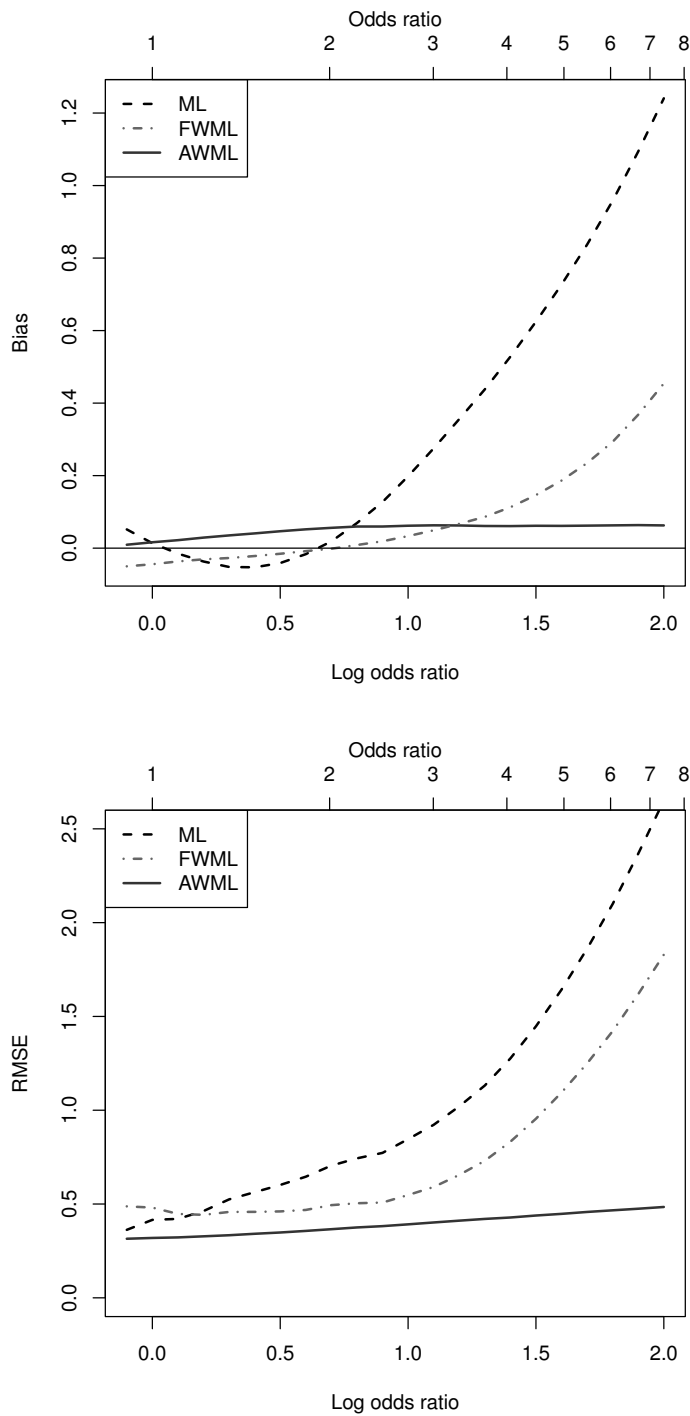}
\end{center}

\label{fig:bias}  
\end{figure}

\begin{figure}
\caption{Simulated coverage probability of the proposed 97.5\% confidence interval.}
\begin{center}
\includegraphics[width=9cm]{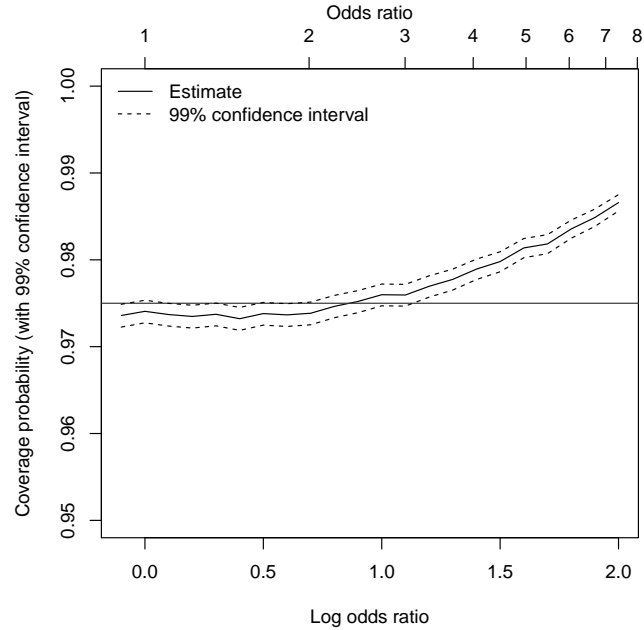}
\label{fig:coverage}
\end{center}
\end{figure}

 As before, 100\,000 replications per scenario were simulated. Bias and root mean square error (RMSE) were determined for the maximum likelihood (ML), fixed weighted maximum likelihood (FWML), and adaptively weighted maximum likelihood (AWML) estimator, while the coverage probability of the proposed confidence interval was evaluated. The weight for the FWML estimator was chosen to be the prespecified weight for the test statistics, i.e. $\omega=w_\I^2=0.5.$ The results for bias and RMSE are shown in Figure \ref{fig:bias}, while the coverage probability is displayed in Figure \ref{fig:coverage}. 

As can be seen in Figure \ref{fig:bias}, the maximum likelihood estimator  is strongly biased  for an increasing treatment effect and is thus not recommended as an estimator for the proposed design. The FWML estimator yields only a small bias for odds ratios less than 3, but is subject to a strong upward bias for larger log odds ratios. The AWML estimator only has a small upward bias with a maximal bias of 0.06 and remains very stable for all considered log odds ratios.
Concerning the RMSE, the AWML estimator shows the uniformly best performance as compared to the two other estimators. The RMSE remains more or less constant for the AWML estimator, while it increases drastically for the other two estimators with increasing treatment effect.

Figure \ref{fig:coverage} shows that the coverage probability of the proposed confidence interval is close to 0.975, and shows a slight upwards tendency with an increasing treatment effect. For log odds ratios smaller than 0.9, the estimated coverage probability was slightly below 0.975, but deviations were small and might also be caused by a sampling error. Overall, the proposed confidence interval seems to perform well in terms of coverage probability.

\section{Discussion}
\label{sec:Discussion}
In this article, we developed and evaluated an adaptive two-stage design which can be used in the context of phase II trials with a large dataset of historical controls at hand. The proposed design overcomes commonly known disadvantages of standard phase II designs in the case that confounding variables are strongly associated with the primary outcome. Our frequentist approach combines two well-estab-lished methodological tools, namely matching of historical controls and adaptive sample size recalculation within a two-stage design, and is novel in two ways. On the one hand, it allows to incorporate matched historical controls within a two-stage single-arm trial. On the other hand, applying the design enables to deal with the uncertainty about trial parameters by means of an interim sample size reassessment. 

In a simulation study, we  first investigated the potential problems of standard phase II designs in case of strong prognostic confounders associated with the outcome. We observed an inflation in the type I error rate for the single-arm design together with a decreased power . For the randomized controlled trial design, there was only a negligible inflation of the type I error rate, together with a slightly decreased power. Subsequently, we demonstrated that our design performs well in terms of maintaining the nominal significance level . Furthermore, it achieves the aspired power for most considered situations and shows a good performance in terms of expected sample size. Regarding  the choice of the effect estimate on which the sample size recalculation is based, using the originally assumed effect yielded a better performance in case this effect was also the true one. Possibly, the approach using the interim estimate  overestimated the true treatment effect in some scenarios yielding a too small sample size leading to an acceptance of $H_0$. In other scenarios, the interim estimate  underestimated the actual treatment effect yielding too large sample sizes. Deviations of the true effect from the assumed one could not be handled sufficiently by using the estimated effect for sample size recalculation. The reason for this issue might be the relatively small number of intervention patients based on which our treatment effect was estimated causing a relatively large standard error for the interim estimate. Hence, it might be sensible to rely on the originally assumed treatment effect when recalculating the sample size. Alternatively, one possible method to overcome this issue might be the use of a combination of interim effect estimate and originally assumed effect, e.g. by using Bayesian Posterior Mean \citep[p.180]{Wassmer2016}.

In case of strong additional variation induced by a residual in the simulation study, the proposed design did not achieve the aspired power values anymore. While the loss in power was rather small in case of a moderate residual standard deviation of $\sigma=0.5$, for a large residual standard deviation of $\sigma=1$, the design only achieved power values between 65\% and 70\%. Unfortunately, the degree of unobserved variation cannot be directly estimated using logistic regression models (unlike in the case of a linear regression model).  However, when the originally planned treatment effect is used for sample size recalculation, our design still yields expected sample sizes which are smaller than the sample size of a comparable fixed design achieving the same power. We also found that the problem of strong residual variance substantially impacts the power of a randomized controlled trial, the loss in power thus also representing a problem for standard designs.

Another crucial aspect of the proposed design is the choice of the futility stopping threshold. We provided approximate formulae for the probability for a futility stop, thus enabling researchers to choose a threshold which performs adequately under both $H_0$ and $H_1$ when planning a trial with our proposed design. Alternatively, the decision to stop for futility can of course also be done using \textit{p}-values or conditional power arguments.

Regarding the choice of suitable control patients, we relied on an iterative procedure and used propensity score matching. Our proposed approach performed well in our simulations. Regarding the choice of the tolerance criterion for the matching procedure, $\tau=0.05$ seemed a good choice in the investigated scenarios. It should be noted that we assumed an equal confounder distribution in intervention and control patients, which might not always be the case in clinical practice. Nevertheless, if this assumption should not be met, our proposed design is expected to perform well nonetheless since differences in terms of confounder distribution between intervention and control patients would merely result in a smaller number of adequate matching partners for each intervention patient. Since our sample size recalculation procedure takes the number of matching partners into account, this situation is expected to solely require larger trial sample sizes but maintaining adequate performance in terms of type I error rate and power.

When choosing a point estimator for the proposed design, the AWML estimator appeared to be the overall best choice in our simulations since it had the overall smallest bias and RMSE when compared to the alternative ML and FWML estimators. The repeated confidence interval for the inverse normal method seems to be a suitable interval estimator since it maintained the nominal coverage probability and has the nice property that it is centered around the well-performing AWML estimator on the log odds ratio scale.

It is a crucial aspect that the historical control data are of high quality, e.g. covariates used for matching should be measured in a standardized way and no missing values should occur for either outcome or covariates. Our approach furthermore makes the implicit assumption that the patients enrolled into the trial are comparable to their counterparts in the historical control cohort. If this cannot be generally assumed, then quality control of the existing control cases should be conducted, e.g. by removing historical control patients which are not comparable to current patients. This will of course result in a smaller pool of control patients and thus increase the required number of intervention patients, but will decrease the potential for bias.

A limitation of our proposed design is that it is not possible to assess its performance analytically. Instead, we had to rely on simulations. In order to achieve a high precision for our estimates of type I error rate and power, we simulated 100\,000 studies, which took about three to four hours per scenario. Of course, increasing the sample size of control or intervention patients, or taking more confounders into account will increase the computational effort. Parallel programming can be used to simulate several scenarios simultaneously though, thus reducing computation time. 

When assessing the treatment effect, we relied on  logistic regression models adjusting for the confounders. The proposed design does however not rely on parametric assumptions, and one might also consider non-parametric tests only adjusting for the matching strata instead of the confounders, e.g. by using the Cochran-Mantel-Haenszel test \citep[p. 114f]{Agresti2002} for hypothesis testing and the associated common odds ratio as interim decision criterion. This method can be straightforwardly implemented in our proposed design, and is also included as an optional approach in our R simulation code.

Our proposed design represents a novel frequentist approach when standard designs  are likely to not maintain the aspired significance level and power due to strong prognostic factors. We are aware of its methodological complexity and the plethora of tuning parameters that need to be chosen adequately and with care. Therefore, we provided our simulation code such that research teams can perform simulation studies in the planning stage of their trial. This will hopefully help to facilitate the use of our design and pave its way into clinical practice. Currently, we are implementing this design in an actual phase II study in AML patients with similar design specifications as shown in Section \ref{sec:Simulation}. We are hoping that the actual implementation of our methods will help to gain visibility for our method and encourage other researchers to use our proposed design as well.

   We acknowledge the fact that the degree of evidence of our design is not comparable with that of a large randomized controlled trial, since our approach only enables to balance known confounders across treatment arms. Unrecognized confounders might still be unequally distributed across the two treatment groups of our design and possibly cause a bias and loss in power \citep{Betensky2002}. Moreover, the proposed design is not immune to a selection bias which might occur when either sampling the historical controls or the intervention patients, but this is also an issue in classical single-arm designs which are still frequently used in phase II oncology trials. We hence encourage the use of our design in the context of the identification of a response signal in phase II trials in order to better inform subsequent randomized phase III trials. This might reduce the number of failed studies and increase the number of successful phase III trials, thus saving resources from the perspective of patients, pharmaceutical companies, and health authorities.

\bigskip
\begin{center}
{\large\bf  DECLARATION OF INTERESTS STATEMENT}
\end{center}

There are no potential conflicts of interest to declare.

\bigskip

\bibliographystyle{Chicago}

\bibliography{Bibliography-MTC}
\newpage
\section*{Appendix: Additional results on type I error rate of the proposed design for further residual standard deviations}

\renewcommand{\thetable}{A.\arabic{table}}
 \setcounter{table}{0}
\begin{table}
\caption {{Simulated type I error rate (probability to reject the null hypothesis $p_{\mathrm{reject}H_0}$), simulated probability for futility stop $p_{\mathrm{stop}}$, calculated approximate probability for a futility stop $\hat{p}_{\mathrm{stop}}$, and expected sample size $\mathrm E[n_\I+\tilde{n}_\II]$ for varying sample sizes $n_\I$ and $n_C$, for $\theta=0$ and residual standard deviation $\sigma=0.5$.}}
\begin{center}
\begin{tabular}{ccccccc}
$\theta_\mathrm{recalc}$ &$n_C$ & $n_\I$ & $p_{\mathrm{reject}H_0}$ &  $p_{\mathrm{stop}}$ & $\hat{p}_{\mathrm{stop}}$ & $\mathrm E[n_\I+\tilde{n}_\II]$ \\
\hline
$\theta_\mathrm{plan}$& 500 		& 20 & 0.0246    & 0.6839   &   0.6868     &  41.91   \\
										& 	500			& 25 & 0.0239    & 0.7061   &   0.7068     & 43.48  \\
										& 	500			& 30 & 0.0243    & 0.7221   &   0.7242     & 46.14  \\
										&   1000	  & 20 &  0.0241   & 0.6930   &   0.6946     &  39.44  \\
										&		1000		&	25 & 0.0236    & 0.7134   &   0.7152     & 41.42  \\
										&		1000		& 30 & 0.0237    & 0.7303  &   0.7333     & 44.29  \\

\hline
$\hat{\theta}_\I		$& 500 		& 20 	 & 0.0244 & 0.6839  &   0.6868      & 41.94  \\
										& 	500			& 25 & 0.0246 & 0.7061 &   0.7068      & 44.35  \\
										& 	500			& 30 & 0.0242 & 0.7221 &   0.7242      & 47.34 \\
										&   1000		& 20 & 0.0239 & 0.6930  &   0.6946      & 40.97   \\
										&		1000		&	25 & 0.0235 & 0.7134  &   0.7152      & 43.65  \\
										&		1000		& 30 & 0.0233 & 0.7303  &   0.7333      & 46.69  \\

\hline
\end{tabular}

\end{center}

\small \textbf{Note}: A matching rate tolerance of $\tau=0.05$ and a residual standard deviation of $\sigma=0.5$ were used to simulate all scenarios in this table.
\label{H0_sigma_0_5} \end{table}

\begin{table}
\caption {{Simulated type I error rate (probability to reject the null hypothesis $p_{\mathrm{reject}H_0}$), simulated probability for futility stop $p_{\mathrm{stop}}$, calculated approximate probability for a futility stop $\hat{p}_{\mathrm{stop}}$, and expected sample size $\mathrm E[n_\I+\tilde{n}_\II]$ for varying sample sizes $n_\I$ and $n_C$, for $\theta=0$ and residual standard deviation $\sigma=1$.}}
\begin{center}
\begin{tabular}{ccccccc}
$\theta_\mathrm{recalc}$ &$n_C$ & $n_\I$ & $p_{\mathrm{reject}H_0}$ &  $p_{\mathrm{stop}}$ & $\hat{p}_{\mathrm{stop}}$ & $\mathrm E[n_\I+\tilde{n}_\II]$ \\
\hline
$\theta_\mathrm{plan}$& 500 		& 20 & 0.0242    & 0.6881   &   0.6868     &  41.30   \\
										& 	500			& 25 & 0.0237    & 0.7082   &   0.7068     & 43.06  \\
										& 	500			& 30 & 0.0235    & 0.7256   &   0.7242     & 45.61  \\
										&   1000	  & 20 &  0.0236   & 0.6981   &   0.6946     &  38.65  \\
										&		1000		&	25 & 0.0237    & 0.7175   &   0.7152     & 40.76  \\
										&		1000		& 30 & 0.0231    & 0.7346   &   0.7333     & 43.64 \\

\hline
$\hat{\theta}_\I		$& 500 		& 20 	 & 0.0238 & 0.6881  &   0.6868      & 41.62   \\
										& 	500			& 25 & 0.0234 & 0.7082  &   0.7068      & 44.25   \\
										& 	500			& 30 & 0.0233 & 0.7256  &   0.7242      & 47.10   \\
										&   1000		& 20 & 0.0237 & 0.6981  &   0.6946      & 40.60    \\
										&		1000		&	25 & 0.0234 & 0.7175 &   0.7152      & 43.37  \\
										&		1000		& 30 & 0.0231 & 0.7346  &   0.7333      & 46.40   \\

\hline

\end{tabular}

\end{center}

\small \textbf{Note}: A matching rate tolerance of $\tau=0.05$ and a residual standard
 deviation of $\sigma=1$ were used to simulate all scenarios in this table.
\label{H0_sigma_1} \end{table}

\end{document}